%% file: main.tex
\pdfoutput=1

\documentclass[11pt]{article}

\usepackage[]{acl}

\usepackage{graphicx}
\usepackage{subfigure}
\usepackage{booktabs} %
\usepackage{listings}
\usepackage{multirow} 
\usepackage{tcolorbox}
\usepackage{colortbl}
\usepackage{times}
\usepackage{latexsym}

\usepackage[T1]{fontenc}
\usepackage{hyperref}
\usepackage{pifont}
\usepackage{subfigure}

\usepackage{xcolor}
\definecolor{lightblue}{RGB}{212,220,247}
\definecolor{lightred}{RGB}{241,208,207}
\definecolor{darkgreen}{RGB}{50,100,0}
\usepackage{amsmath}
\usepackage{amssymb}
\usepackage{mathtools}
\usepackage{amsthm}
\usepackage{fontawesome5}
\usepackage{xspace}

\usepackage[capitalize,noabbrev]{cleveref}

\theoremstyle{plain}

\theoremstyle{definition}

\theoremstyle{remark}

\usepackage[textsize=tiny]{todonotes}
\newcommand{\yes}{\color{darkgreen}\ding{52}}
\newcommand{\no}{\color{red}\ding{56}}

\usepackage{microtype}

\usepackage{inconsolata}
\usepackage{amsmath}
\usepackage{amssymb}
\usepackage{mathtools}
\usepackage{amsthm}
\usepackage{graphicx}
\usepackage{float}

\title{HumanEval Pro and MBPP Pro:  Evaluating Large Language Models \\ on Self-invoking Code Generation}

\author{Zhaojian Yu$^1$ \quad Yilun Zhao$^2$ \quad Arman Cohan$^{2}$ \quad Xiao-Ping Zhang$^1$ \vspace{4pt}\\
$^1$Tsinghua University \quad $^2$Yale University 
}

\addtocontents{toc}{\protect\setcounter{tocdepth}{-1}}

\begin{document}
\maketitle

\begin{minipage}[t]{2\linewidth}
\vspace{-1.75cm}
  \centering
  \href{https://github.com/CodeEval-Pro/CodeEval-Pro}{{\faGithub{}}~~\texttt{\small github.com/CodeEval-Pro/CodeEval-Pro}}
\vspace{0.5cm}
\end{minipage}

\begin{abstract}
We introduce self-invoking code generation, a new task designed to evaluate the progressive reasoning and problem-solving capabilities of LLMs. In this task, models are presented with a base problem and a related, more complex problem. They must solve the base problem and then utilize its solution to address the more complex one.
This work features three key contributions.
First, we propose a general recipe for generating more challenging versions of existing benchmarks, resulting in three new benchmarks: HumanEval Pro, MBPP Pro, and BigCodeBench-Lite Pro, specifically designed to assess LLMs on self-invoking code generation.
Second, from the analysis of experimental results over twenty LLMs on our benchmarks, we have two important observations: (i) Most LLMs excel in traditional code generation benchmarks like HumanEval and MBPP, but their performance declines on self-invoking tasks. For example, o1-mini achieves 96.2\% pass@1 on HumanEval but only 76.2\% on HumanEval Pro. (ii) On self-invoking code generation task, the  instruction-tuned models demonstrate only marginal improvements compared to the base models. Third, we disclose the types of failure modes that exist in our evaluation results. All these results underscore the need for further advancements in self-invoking code generation tasks and provide a new 
direction for future research on enhancing LLMs' code reasoning capabilities.
\end{abstract}

\begin{figure}[t]
    \centering
    \includegraphics[width=\linewidth]{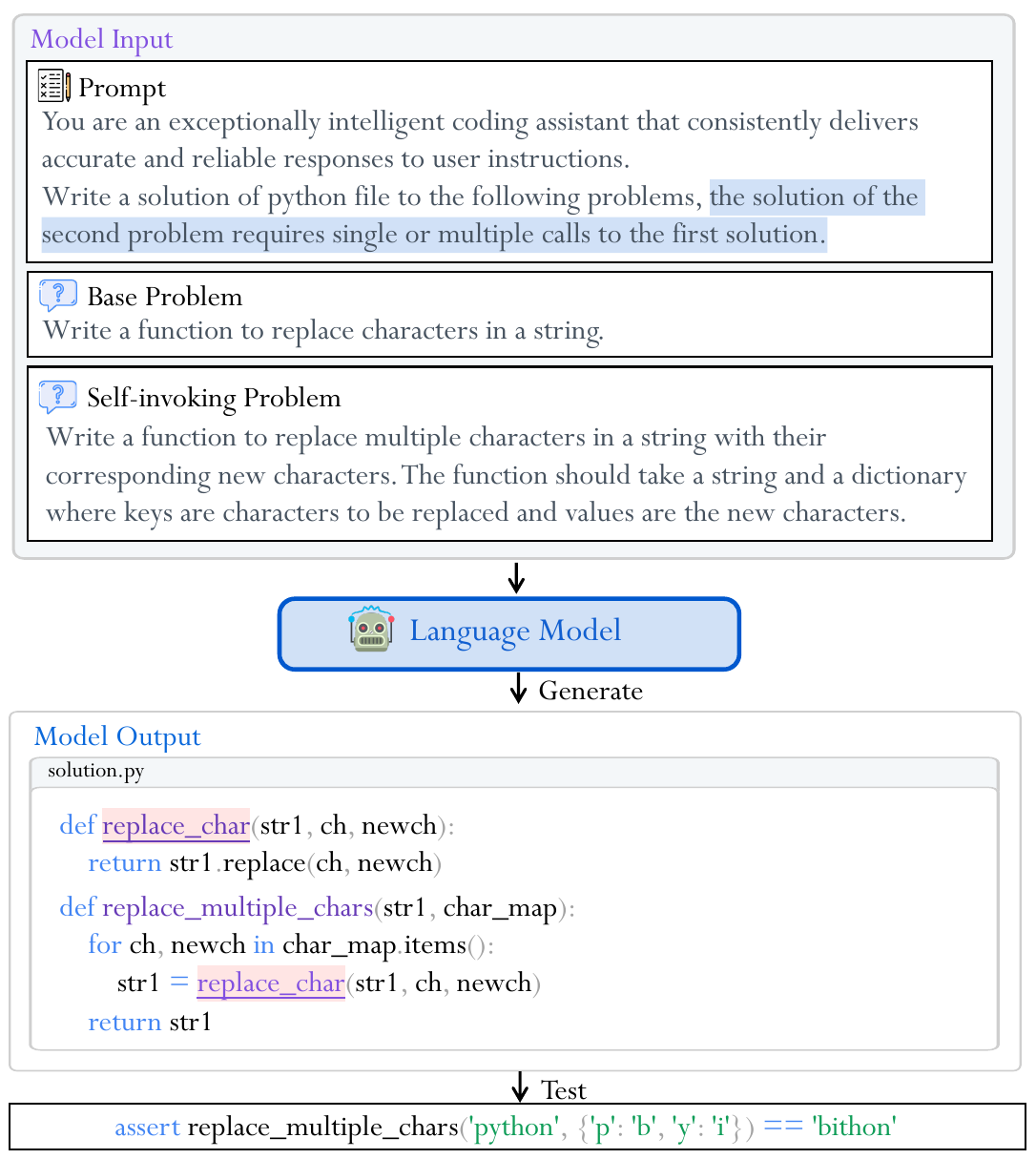}
    \caption{The overview of self-invoking code generation in HumanEval Pro and MBPP Pro. Given a base problem and a related, more complex problem, they are required to solve the base problem and use its solution to address the complex problems.}
    \label{fig:overview}
\end{figure}

\section{Introduction}
\label{sec:intro}
Large Language Models (LLMs) have demonstrated significant progress in various code-related tasks including code generation \cite{roziere2023code, zhang2023survey, ni-etal-2024-l2ceval}, program repair \cite{xia2022practical, jin2023inferfix}, and code translation \cite{zhu2022xlcost}, etc. 
Traditional human-annotated benchmarks such as HumanEval \cite{chen2021evaluating} and MBPP \cite{austin2021program} have been widely adopted to evaluate the code generation abilities of LLMs, providing standardized evaluation protocols for assessing their performance on code-related tasks. However, these existing benchmarks primarily focus on isolated, single-function code generation, which represents only a subset of the challenges encountered in real-world software development scenarios.

To evaluate LLMs under more realistic problem-solving scenarios, 
BigCodeBench \cite{zhuo2024bigcodebench} presents a benchmark that comprises of complex and practical problems requiring LLMs to use multiple function calls from diverse libraries. While BigCodeBench highlights the use of \emph{external} function calls, it falls short in assessing LLMs' reasoning ability to generate and invoke their own generated functions in problem-solving. CRUXEval \cite{gucruxeval} assesses LLMs' code reasoning by predicting function inputs and outputs. However, the direct input and output prediction does not involve explicit code generation.
In practical software engineering contexts, developers must not only write code but also comprehend, modify, and utilize existing code to solve more complex problems. 
Hence, the ability to understand and subsequently leverage one's own generated code, namely \emph{self-invoking code generation} (\Cref{fig:overview}), plays an important role for LLMs to leverage their reasoning capabilities to code generation that current benchmarks fail to capture.

Therefore, we present \textbf{HumanEval Pro} and \textbf{MBPP Pro}, two expanded versions of the traditional HumanEval and MBPP benchmarks to evaluate LLMs on self-invoking code generation task.
As illustrated in \cref{fig:overview}, HumanEval Pro and MBPP Pro extend beyond simple code generation by introducing self-invoking problems which requires LLMs to solve the base problem and invoke their self-generated code to solve a more complex problem.
By evaluating LLMs on self-invoking code generation task, HumanEval Pro and MBPP Pro provide a useful and important probe to better understand the programming capabilities of LLMs. The capability of self-invoking code generation also facilitates LLMs to tackle difficult tasks with greater autonomy and effectiveness. 

To obtain HumanEval Pro and MBPP Pro, we propose a general recipe for constructing self-invoking code generation benchmarks by building upon existing datasets.
First, we use Deepseek-V2.5~\cite{deepseekv2} to generate self-invoking problems based on the original problems in HumanEval and MBPP. These problems are designed to be more complex than the base problems and closely related to them, ensuring progressive reasoning and coherent code invocation.
Second, we generate the candidate solution and  test inputs for each problem.
Third, we execute the code of candidate solution to generate output and use the \textit{assert} command in Python to build test cases. In the third stage, human experts are assigned to manually review each problem and continuously modify and execute the code of solutions to ensure that all canonical solutions could correctly solve the problem and cover the test cases. 
To verify the reproducibility of our benchmark construction approach, we further construct BigCodeBench-Lite Pro, a new self-invoking problems set derived from  BigCodeBench \cite{zhuo2024bigcodebench}. On Bigcodebench-Lite Pro, LLMs show consistent performance trend with HumanEval Pro and MBPP Pro, which emphasizes the generalizability of our construction pipeline.
Therefore, our benchmark construction approach can also be extended to adapt other code generation benchmarks, particularly as the capabilities of LLMs advance and older benchmarks become obsolete.

Through extensive evaluation of various LLMs, we uncover a significant disparity between traditional code generation and self-invoking code generation capabilities. Our findings reveal that while frontier LLMs excel at generating individual code snippets, they often struggle to effectively utilizing their own generated code for solving more complex problems.
For example, o1-mini achieves
96.2\% pass@1 on HumanEval but only 76.2\%
on HumanEval Pro, demonstrating the  challenges inherent in self-invoking code generation.
From the comparison between instruction-tuned models and their base models, we found that instruction-tuned models are less efficient on self-invoking code generation than traditional code generation task.
Furthermore, 
our detailed statistics of failure cases in HumanEval Pro and MBPP Pro also reflect the shortcomings of LLMs in self-invoking code generation, thereby providing complementary insights on real-world coding capabilities of LLMs.

\begin{figure*}[t!]
    \centering
    \includegraphics[width=\linewidth]{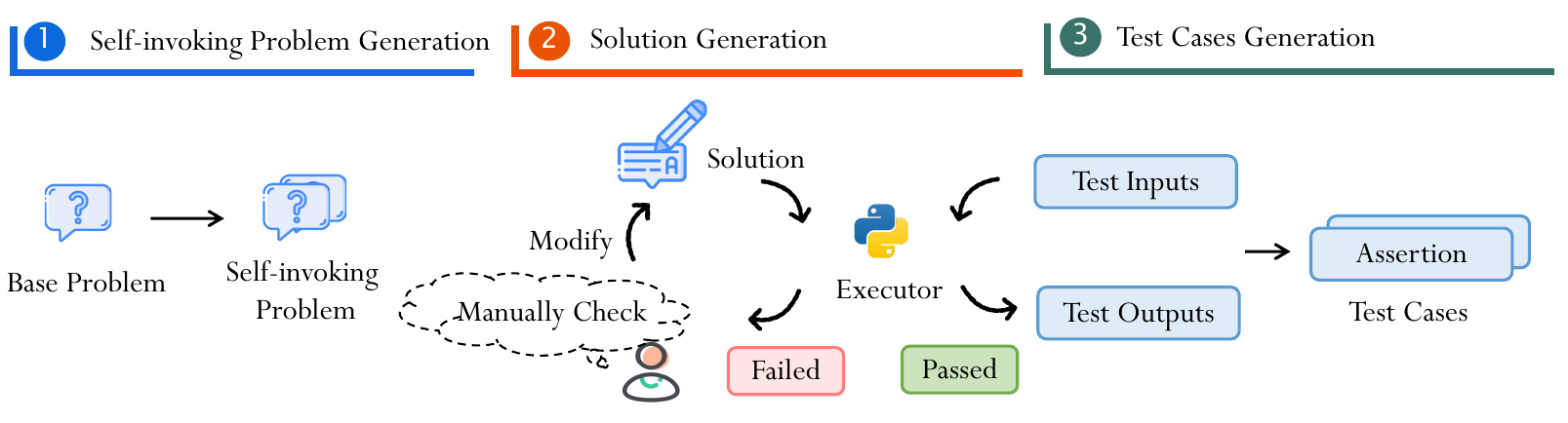}
    \caption{The overview of benchmark construction. An example is shown in \cref{fig:bc_exp}. We summarize the entire benchmark construction process as follows:
(1) \textbf{Self-invoking problem Generation}: We use Deepseek-V2.5 to generate the self-invoking problems, as well as their candidate solutions and test inputs.
(2) \textbf{Solutions Generation}: We execute the generated solution with the test inputs in a controlled Python environment to obtain ground truth outputs.
(3) \textbf{Test Cases Generation}: 
We employ an iterative method involving Python execution check and manual review to ensure that all test cases pass successfully.
The final execution results are then used to construct complete test cases with \texttt{assert} command.}
    \label{fig:benchmark-construction}
\end{figure*}

\section{Related Work}
Recent advances in LLMs have demonstrated remarkable capabilities in code generation and understanding. This section reviews the current landscape of code-related benchmarks and LLMs.

\paragraph{Benchmarks for Code Generation}
The evaluation landscape for Code LLMs has evolved significantly. HumanEval \cite{chen2021evaluating} and MBPP \cite{austin2021program} serve as fundamental benchmarks, focusing on Python function completion tasks with test-driven evaluation. 
Several benchmarks have expanded code evaluation benchmarks to encompass multiple programming languages~\cite{zheng2023codegeex, athiwaratkun2022multi}, complex tasks like program repair~\cite{haque2022fixeval, jiang2023impact,muennighoff2024octopack, evoeval}, dynamic problem sets~\cite{jain2024livecodebench}, and code reasoning through code summarization \cite{barone2017parallel, hasan2021codesc} and simulated execution~\cite{gucruxeval}.
To evaluate LLMs in professional software engineering, benchmarks like SWE-Bench~\cite{jimenez2023swe}, EvoCodeBench~\cite{li2024evocodebench}, RepoBench~\cite{liu2023repobench}, and GoogleCodeRepo~\cite{shrivastava2023repository} focus on real-world tasks, code evolution, and repository-level challenges.
These benchmarks collectively drive the advancement of LLMs, 
providing valuable insights into their strengths and limitations. Our benchmarks introduce novel self-invoking code generation task, which addresses gaps left by existing benchmarks. This addition provides a more holistic framework to evaluate LLMs on leveraging their
reasoning capabilities to code generation.
Moreover, our benchmark construction method could also push existing benchmarks forward to accommodate more complex and challenging code-related tasks.

\vspace{-0.2cm}
\paragraph{LLMs for Code Generation}
The development of foundation models specifically designed for code generation has seen significant progress. 
CodeX \cite{chen2021evaluating} pioneered this direction by fine-tuning GPT models on code-specific data. Subsequent models like CodeGeeX \cite{zheng2023codegeex} and CodeLLaMA \cite{roziere2023code} further advanced the field by incorporating multilingual code understanding and generation capabilities. StarCoder \cite{li2023starcoder}, DeepseekCoder \cite{zhu2024deepseek2} and Qwen2.5-Coder \cite{hui2024qwen2} demonstrated the importance of high-quality code data curation and specialized architecture designs.
Building upon these models, researchers have explored instruction-tuning approaches using GPT-4 or GPT-3.5 as teachers. Notable examples include WizardCoder \cite{luo2023wizardcoder},  Magicoder \cite{wei2024magicoder}, WaveCoder \cite{yu2024wavecoder}, OpenCodeInterpreter \cite{zheng2024opencodeinterpreter}, and ReflectionCoder \cite{ren2024reflectioncoder}. These models have achieved impressive performance on standard code generation benchmarks through enhanced data diversity and instruction complexity.

\section{Benchmark Construction}
\label{sec3:bench}
To facilitate a meaningful comparison between self-invoking code generation and traditional code generation, we have crafted two new benchmarks, HumanEval Pro and MBPP Pro. 
These benchmarks are extensions 
of the original HumanEval and MBPP, requiring the model to solve both the base problem and a more complex self-invoking problem.
In addressing the self-invoking problems, LLMs are required to apply the solutions they have independently generated for the base problem. This evaluation of self-invoking code generation offers deeper insights into the  programming capabilities of LLMs, extending beyond the scope of single-problem code generation.
The benchmark construction process, illustrated in \cref{fig:benchmark-construction}, will be discussed in detail in the following subsections.

\subsection{Self-invoking Problem Generation}
To ensure that all benchmarks are permissively licensed, we employ one of the state-of-the-art (SoTA) open-source models, DeepSeek-V2.5, to create new problems and solutions derived from the original HumanEval and MBPP datasets. Two main guidelines is established for self-invoking problems generation to rigorously evaluate LLMs.
1) \textbf{Complexity Enhancement}: The self-invoking  problems should introduce additional programming challenges while preserving the core functionality of the original problems. This ensures that successful solutions require both understanding of the original code and ability to extend it appropriately.
2) \textbf{Semantic Relevance}: The self-invoking  problems should maintain sufficient semantic similarity to their original counterparts to enable meaningful self-invoking code generation process. \cref{app:bench-prompt} presents the prompt for self-invoking problem generation. 
\subsection{Solution Generation}
In self-invoking problem generation process, the candidate solution and test inputs are generated simultaneously with the self-invoking problem.
However, when dealing with self-invoking problems, these generated solutions are often flawed, which can lead to execution errors during the verification process, thereby highlighting a significant challenge in maintaining the accuracy and effectiveness of these test cases. 
Therefore, as shown in \cref{fig:benchmark-construction}, we propose a method to iteratively execute the solution code with test inputs and obtain expected outputs correctly. 
For the execution errors, the authors manually analyze these errors and modify the solutions to ensure that the final solution can cover all the test cases comprehensively. 
The manual review process involves (1) identifying the root causes of the errors, (2) making necessary adjustments to the code or algorithm, and (3) re-evaluating the solution against the entire set of test cases to confirm its correctness and completeness.
\cref{tab:human-val} shows that our rigorous verification process 
ensures the high quality of our benchmarks.

\subsection{Test Cases Generation}
After obtaining the self-invoking problem and its candidates solution, a critical challenge is ensuring the reliability of the test cases (with both test inputs and expected execution outputs) to validate the the generated solutions.  
Despite the apparent simplicity of using the same LLM context to generate both problems and test cases, CRUXEval \cite{gucruxeval} results show that even leading models like GPT-4 achieve only a 63.4\% pass@1 rate in test output prediction.
This suggests that using models like GPT-4 to directly generate test cases for problems will lead to many inaccurate evaluation results.
Our iterative verification method effectively addresses this challenge. By combining Python execution checks with manual reviews, we ensure that all test cases accurately assess solution correctness and achieves a 100\% pass@1 under correct implementation conditions.
\begin{table}[!t]
\small
\centering

\begin{tabular}{l|cc}
\toprule
\textbf{Iteration} & \textbf{HumanEval Pro (\%)} & \textbf{MBPP Pro (\%)} \\
\midrule
Round 1     & 64.0          & 84.7     \\
Round 2    & 98.8          & 99.7     \\
Round 3     & 100.0           & 100.0      \\
\bottomrule
\end{tabular}
\caption{Pass@1 (\%) of candidate solutions across different iteration rounds for canonical solution and test case generation with human manual review.}
\label{tab:human-val}
\end{table}
Furthermore, we categorize the common execution errors that occur during test case generation
into four main types:
\emph{variable type mismatches},
\emph{index out of bounds},
\emph{invalid input handling},
and \emph{edge case failures}.
To obtain the high-quality self-invoking problem solutions, we adopt main remediation strategies including:
(1) implementing input validation,
(2) adding type checking,
(3) handling edge cases explicitly,
and (4) refining problem specifications when necessary.
Beyond basic execution correctness, we also verify the self-invoking problem and solutions in the following aspects:
(1) logical consistency between problem statements and test cases,
(2) coverage of essential edge cases,
and (3) alignment with original problem objectives.

\begin{table*}[h]
\centering
\small
\renewcommand{\arraystretch}{0.95}

\begin{tabular}{l|c|c|cc|c|cc}
\toprule
\multirow{2}{*}{\textbf{Model}} & \multirow{2}{*}{\textbf{Params}} & \multirow{2}{*}{\textbf{HumanEval (+)}} & \multicolumn{2}{|c|}{\textbf{HumanEval Pro}} & \multirow{2}{*}{\textbf{MBPP (+)}} & \multicolumn{2}{|c}{\textbf{MBPP Pro}} \\
& &  & (0-shot) & (1-shot) &  & (0-shot) & (1-shot)  \\
\midrule
\multicolumn{8}{c}{Proprietary Models} \\
\midrule
o1-mini & - & 97.6 (90.2) &  76.2 & 84.8  & 93.9 (78.3) &  68.3 & 81.2 \\
GPT-4o & - & 90.2 (86.0) &  75.0 & 77.4  & 86.8 (72.5) &  70.9 & 80.2\\
GPT-4-Turbo &- &90.2 (86.6) & 72.0 & 76.2 & 85.7 (73.3) & 69.3 & 73.3 \\
Claude-3.5-sonnet & - & 92.1 (86.0) &  72.6 & 79.9  & 91.0 (74.6) &  66.4 & 76.2 \\
\midrule
\multicolumn{8}{c}{Open-source Models} \\
\midrule

Deepseek-V2.5 & - & 90.2 (83.5) & 73.8 & 76.8 & 87.6 (74.1) & 71.2 & 77.5 \\
DeepseekCoder-V2-instruct & 21/236B & 90.2 (84.8) & 77.4 & 82.3 & 89.4 (76.2) & 71.4 & 76.5 \\
\midrule
Qwen2.5-Coder-1.5B-base & 1.5B & 43.9 (36.6) & 37.2 & 39.6 & 69.2 (58.6)  &  48.4 & 51.3 \\
Qwen2.5-Coder-1.5B-instruct & 1.5B & 70.7 (66.5) & 33.5 & 37.8 & 69.2 (59.4) &  42.1 & 43.7 \\
\midrule
DeepseekCoder-6.7B-base & 6.7B & 49.4 (39.6) & 35.4 & 36.6 & 70.2 (51.6) & 50.5 & 55.0 \\
DeepseekCoder-6.7B-instruct & 6.7B & 78.6 (71.3) & 55.5 & 61.6 & 74.9 (65.6) & 57.1 & 58.2 \\
Magicoder-S-DS-6.7B & 6.7B & 76.8 (70.7) &  54.3 & 56.7 & 75.7 (64.4) &  58.7 & 64.6  \\
WaveCoder-Ultra-6.7B & 6.7B & 78.6 (69.5) &  54.9 & 59.8 & 74.9 (63.5) &  60.1 & 64.6 \\
\midrule
Qwen2.5-Coder-7B-base & 7B & 61.6 (53.0) &  54.9 & 56.1 & 76.9 (62.9) & 61.4 & 68.0 \\
Qwen2.5-Coder-7B-instruct & 7B & 88.4 (84.1) & 65.9 & 67.1 & 83.5 (71.7) & 64.8 & 69.8 \\
\midrule
OpenCoder-8B-base & 8B & 66.5 (63.4) & 39.0 & 42.1  & 79.9 (70.4) & 52.4 & 53.7 \\
OpenCoder-8B-instruct & 8B & 83.5 (78.7) & 59.1 & 54.9 & 79.1 (69.0) &  57.9 & 61.4  \\
\midrule
Yi-Coder-9B-base & 9B & 53.7 (46.3) & 42.7 & 50.0  & 78.3 (64.6) & 60.3 & 61.4 \\
Yi-Coder-9B-chat & 9B & 85.4 (74.4) & 59.8 & 64.0 & 81.5 (69.3) & 64.8 & 71.7 \\
\midrule
Codestral-22B-v0.1 & 22B & 81.1 (73.2) & 59.1 & 65.9 & 78.2 (62.2) & 63.8 & 71.2  \\
\midrule
DeepseekCoder-33B-base & 33B & 56.1 (47.6) & 49.4 & 49.4 & 74.2 (60.7) & 59.0 & 65.1 \\
DeepseekCoder-33B-instruct & 33B & 79.3 (75.0) & 56.7 & 62.8 & 80.4 (70.1) & 64.0 & 68.3 \\
\midrule
Qwen2.5-Coder-32B-base & 32B & 65.9 (60.4) & 61.6 & 67.1 & 83.0 (68.2) & 67.7 & 73.3 \\
Qwen2.5-Coder-32B-instruct& 32B & 92.7 (87.2)  & 70.1 & 80.5 & 90.2 (75.1) & 69.8 & 77.5 \\
\midrule
LLaMA3-70B-instruct & 70B & 81.7 (72.0) & 60.4 & 64.6 & 82.3 (69.0) &  63.5 & 70.4 \\
\bottomrule
\end{tabular}
\caption{Main result of different models on HumanEval Pro and MBPP Pro. More results  is shown in \cref{sec:res}. }
\label{tab:result}
\end{table*}

\section{Experiments}
We present results of proprietary models and open-source models on HumanEval Pro and MBPP Pro: Qwen-2.5-Coder (Base and Instruct, 1.5B, 7B, 33B)~\cite{hui2024qwen2},
DeepseekCoder (Base and Instruct)~\cite{deepseek-coder}, DeepseekCoder-V2~\cite{deepseekv2}, Yi-Coder-9B (Base and Instruct)~\cite{yicoder}, OpenCoder (Base and instruct)~\cite{huang2024opencoder}, Magicoder-S-DS-6,7B~\cite{wei2024magicoder}, WaveCoder-Ultra-6.7B~\cite{yu2024wavecoder}, Codestral-22B~\cite{mistral2024codestral}, GPT-3.5~\cite{ouyang2022training}, GPT-4o~\cite{gpt4o},  Claude-3.5-sonnet~\cite{TheC3} and o1-mini~\cite{OpenAIOS}. To facilitate reproducibility, the HuggingFace checkpoints of all open-source models and API name of proprietary models are provided in Appendix \ref{app:model}. Our prompts for evaluation is shown in \cref{app:prompt}.

\begin{figure*}[!t]
    \centering
    \includegraphics[width=\linewidth]{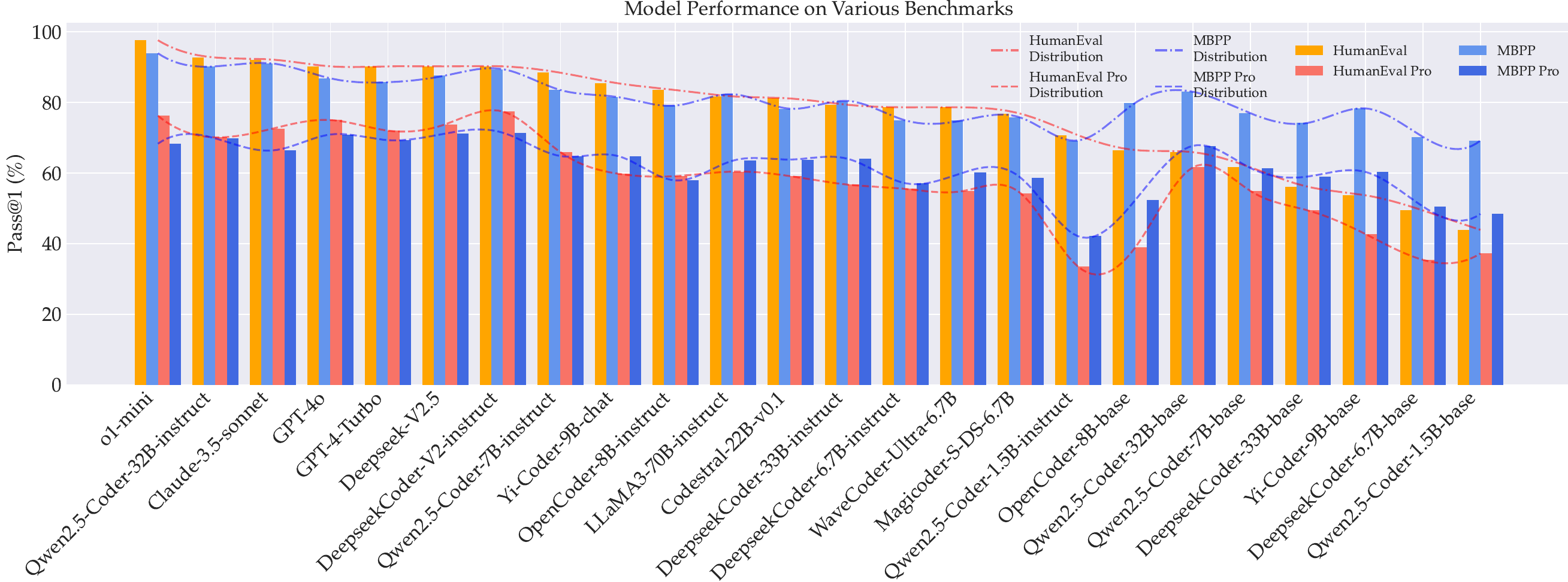}
    \caption{Performance Comparison: HumanEval Pro (and MBPP Pro) vs. HumanEval (and MBPP).}
    \label{fig:bench-comp}
\end{figure*}

Following previous work~\cite{chen2021evaluating}, We use the \textit{pass@k} \cite{chen2021evaluating} score as the evaluation metric of HumanEval Pro and MBPP Pro. We use greedy decoding strategy to generate solutions for all open-source models and set \textit{temperature=0.2} for all API-models.
For all previous benchmarks, we use the reported results whenever available; otherwise, we evaluate using the EvalPlus codebase~\cite{liu2024your}. 

\label{sec:ana}
\begin{figure*}[t]
    \centering
    \includegraphics[width=0.95\linewidth]{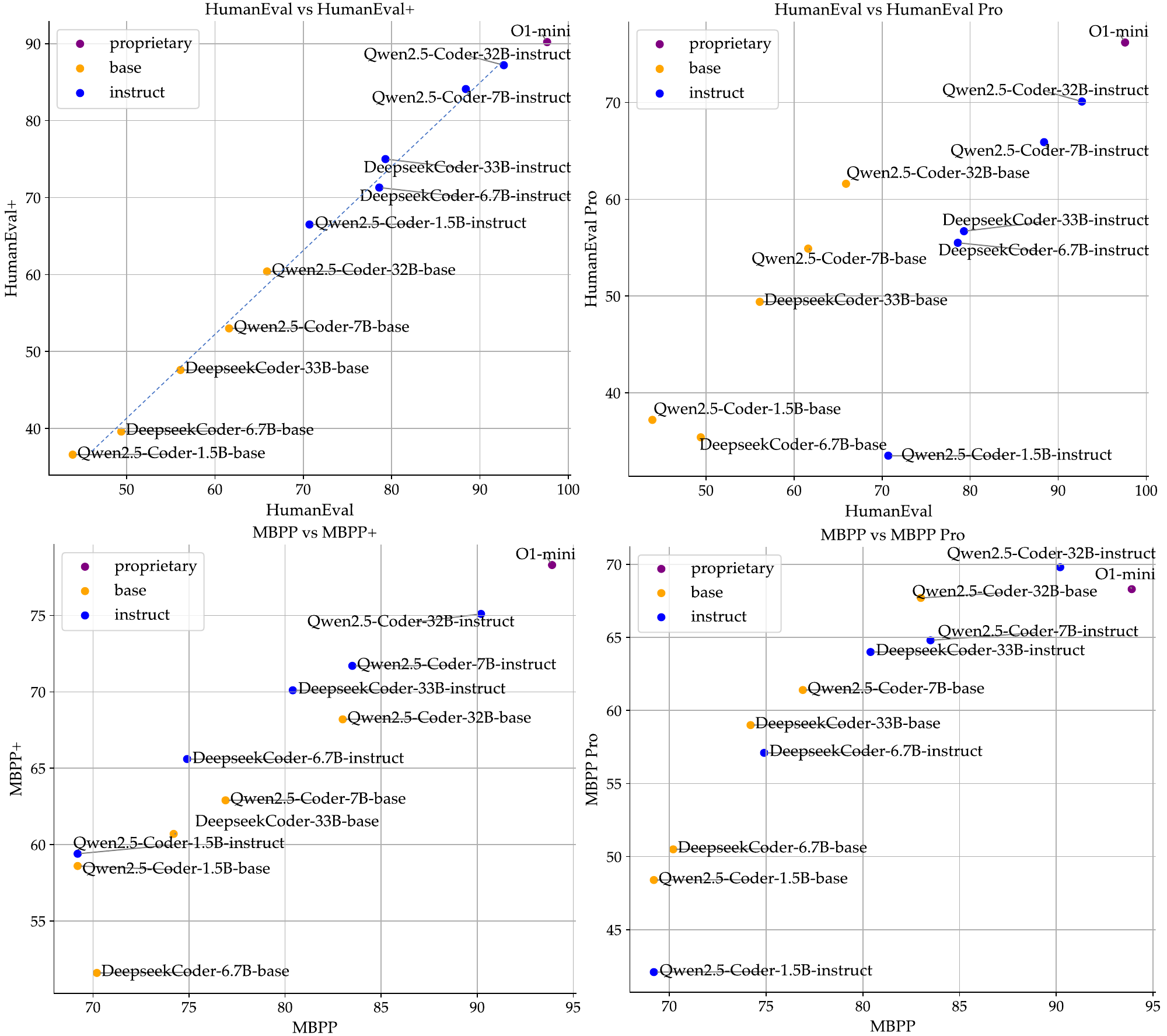}
    \caption{HumanEval (or MBPP) scores against the results on HumanEval Pro and MBPP Pro (HumanEval+ and MBPP+). We presents the comparison between base model and instruct model. 
    }
    \label{fig:comp}
\end{figure*}
\cref{tab:result} presents the \textit{pass@1} scores of HumanEval Pro and MBPP Pro alongside those of other relevant benchmarks, including HumanEval, HumanEval+, MBPP, and MBPP+ \cite{liu2024your}, highlighting the following salient observations:
1) Most LLMs have a 10\% to 15\% absolute performance drop on self-invoking code generation benchmarks.
2) Large size open-source LLMs have comparable performance with proprietary LLMs on self-invoking benchmarks. Notably, DeepseekCoder-V2-instruct achieves 77.4\% on HumanEval Pro, surpassing the score of all proprietary LLMs.
3) Most instruction-tuned models have less improvements on self-invoking code generation benchmarks (e.g., HumanEval Pro) than traditional benchmarks (e.g.,HumanEval). For instance, Qwen2.5Coder-32B-instruct have 26.8\% absolute improvement on HumanEval compared to Qwen2.5Coder-32B-base (from 65.9\% to 92.7\%) but only 8.5\% on HumanEval Pro (from 61.6\% to 70.1\%). 
\cref{sec:res} also presents the evaluation results for different $k$ values with the sampling generation strategy.
\cref{sec:ana} provides detailed analysis for these results.

\section{Analysis}

\begin{figure*}[t]

\centering

\subfigure[Qwen2.5-Coder-7B-base]{\includegraphics[width = 0.49\textwidth]{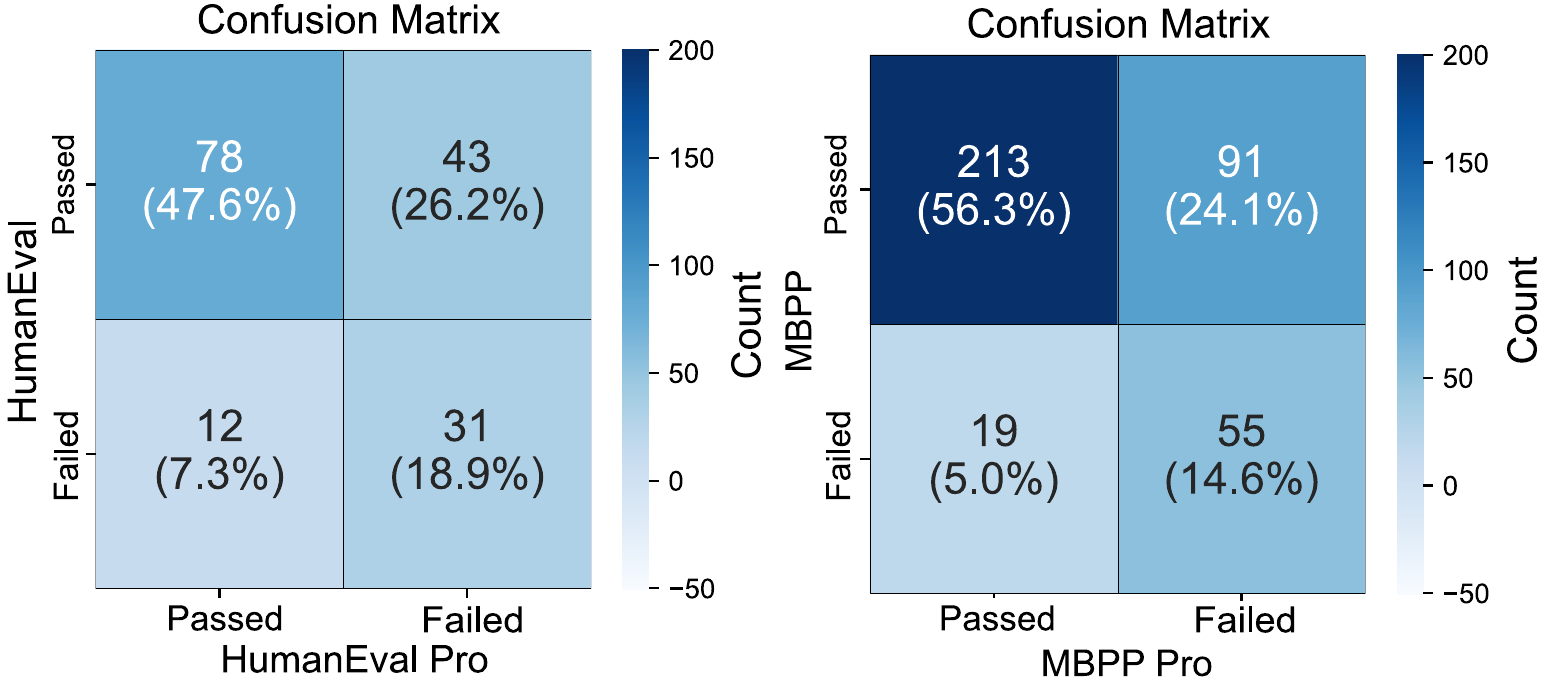}}
\subfigure[Qwen2.5-Coder-32B-base]{\includegraphics[width = 0.49\textwidth]{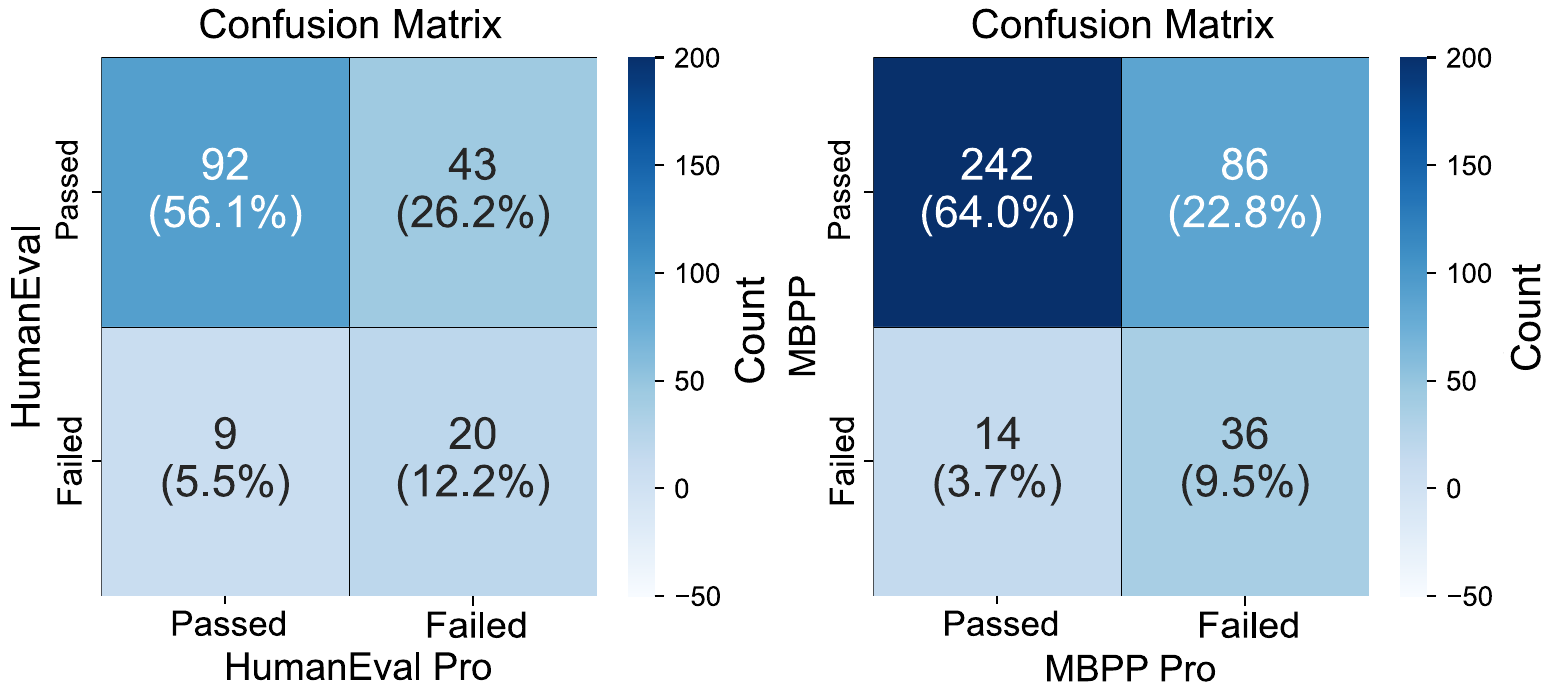}}

\subfigure[Qwen2.5-Coder-7B-instruct]{\includegraphics[width = 0.49\textwidth]{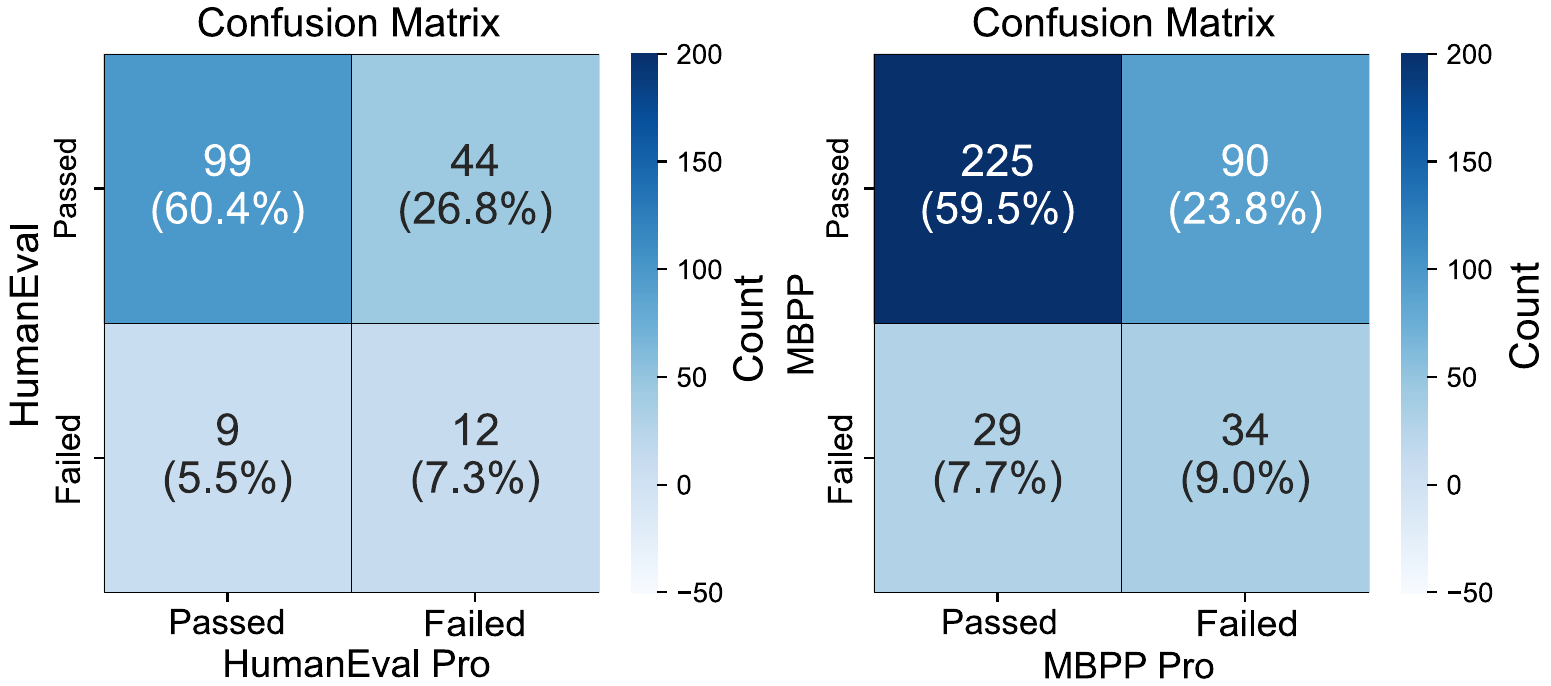}}
\subfigure[Qwen2.5-Coder-32B-instruct]{\includegraphics[width = 0.49\textwidth]{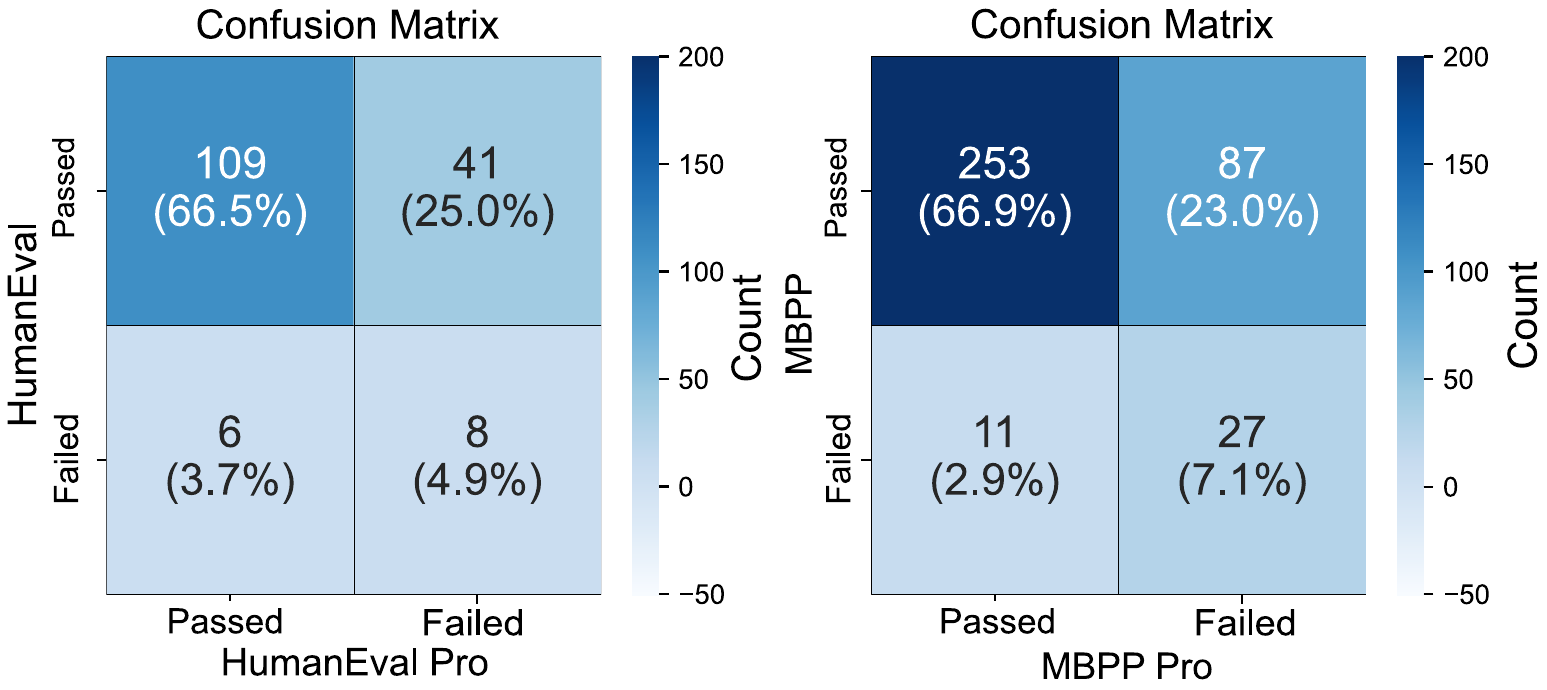}}

\caption{The confusion matrix of different models. We use \texttt{(Failed, Passed)} to indicate samples that fail in HumanEval Pro (or MBPP Pro) but pass in HumanEval (or MBPP).
}
\label{fig:matrix}
\end{figure*}

\textbf{Frontier LLMs still face challenges in self-invoking code generation.} 
\cref{tab:result} and \cref{fig:bench-comp} present the comparison between HumanEval Pro (or MBPP Pro) and HumanEval (or MBPP).
As shown in \cref{tab:result}, while 1-shot prompting improves model performance on HumanEval Pro and MBPP Pro, the pass@1 scores achieved on these datasets remain notably lower compared to their counterparts on the original HumanEval and MBPP benchmarks. This performance gap indicates that although current LLMs excel at direct code generation tasks, they struggle to maintain comparable performance when tasked with self-invoking code generation for complex problems. Notably, even the SoTA reasoning model o1-mini, that achieves an impressive 96.2\% pass@1 on HumanEval, demonstrates significant performance degradation when tackling more complex problems, as evidenced by its lower 76.2 pass@1 score on HumanEval Pro under zero-shot setting.  
\subsection{Base Model vs Instruct Model}
\label{sec:comp}
Currently, the training of LLMs is typically divided into two stages: a pre-training stage that relies on self-supervised learning, and a subsequent supervised fine-tuning stage based on $<$instruction, response$>$ pairs. Previous studies~\cite{luo2023wizardcoder, hui2024qwen2, wei2024magicoder} have shown that the instruction-based supervised fine-tuning stage can significantly enhance the code generation capabilities of base models on traditional benchmarks.
For example, as shown in \cref{tab:result}, Qwen2.5-Coder-instruct 7B started with the Qwen2.5-Coder-7B base model and improved the HumanEval pass@1 score from 61.6\% to 88.4\%.
There remains new curiosity about whether these instruction-tuned models still show such significant improvements under a new problem solving scenario. In this section, we explore this through our new benchmarks.

\textbf{The instruction-tuned models demonstrate only marginal improvements compared to the base models on self-invoking code generation. }
In \cref{fig:comp}, we plot the previous reported HumanEval (or MBPP) scores against the results on HumanEval Pro and MBPP Pro (HumanEval+ and MBPP+). 
From the \cref{fig:comp}, we have an interesting finding:  When observing the correlation between HumanEval (or MBPP) and HumanEval Pro (or MBPP Pro), we see that the orange dot (indicates base model) is always to the upper left of the blue dot (indicates instruction-tuned model). 
However, for the comparison between HumanEval (or MBPP) and HumanEval+ (or MBPP+), the blue dot is always distributed to the upper of orange dot (even in a line on HumanEval vs HumanEval+).
Overall, this suggests that while instruction-based fine-tuning significantly improves performance on simpler benchmarks like HumanEval (+) (or MBPP (+)), its efficiency diminishes for more complex self-invoking code generation tasks.  On the other hand, base models like Qwen2.5-Coder-base and Deepseek-Coder-base have a higher  
\begin{equation}
\small
\text{Ratio} = \frac{\text{pass@k on HumanEval Pro (or MBPP Pro)}}{\text{pass@k on HumanEval (or MBPP)}}
\end{equation}
than instruct models, which indicates that they have elevated training potential on self-invoking code generation task.

\begin{table*}[t]
\small
\centering

\begin{tabular}{ll|cc}
\toprule
\textbf{Error Type} & \textbf{Description} &  \textbf{Examples} \\

\midrule
AssertionError      &  Failing to pass the test cases.        & Examples in \cref{app:assertionerror}            \\
NameError           & The code includes undefined variables.   & Examples in \cref{app:nameerror}             \\
ValueError          & Unaware of the value of variables        & Examples in \cref{app:valueerror}              \\
IndexError           & Array out of bounds                    & Examples in \cref{app:indexerror}             \\
TypeError          & Incorrect variable type usage.          & Examples in \cref{app:typeerror}             \\
Other Errors        &  KeyError, SyntaxError, ZeroDivisionError, IndentationError, etc. & -- \\
\bottomrule
\end{tabular}
\caption{The execution error types and their descriptions in our evaluation results.}
\label{tab:error-type}
\end{table*}
\subsection{Confusion Matrix Correlation for Different Models}
From \cref{tab:result}, we observe that most LLMs have a score gap between direct code generation and self-invoking code generation tasks. To better understand the correlation and overlap between these two kinds of tasks, we compare the number of problems passed and failed in HumanEval Pro and MBPP Pro with their corresponding base problems in HumanEval and MBPP.  \cref{fig:matrix} presents an array of confusion matrix over problems, highlighting the following salient observations:

\textbf{Most LLMs are proficient in code generation tasks but struggle with generating code that can self-invoke effectively.}
Although some SoTA LLMs such as Qwen2.5-Coder-32B-instruct successfully solve 90\% of base problems on the original HumanEval and MBPP benchmarks, over 25\% of problems still fail on more challenging HumanEval Pro and MBPP Pro benchmarks with self-invoking code generation (as shown in the top right of each subfigure in \cref{fig:matrix}). This  suggests that the drop in the model's scores on HumanEval Pro and MBPP Pro is largely due to its lower accuracy in generating self-invoking code compared to direct code generation.

\textbf{The instruction-tuned model does not significantly outperform the base model in self-invoking code generation task.} From the confusion matrices of the base model and the instruct model in \cref{fig:matrix}, we can observe a trend: the instruction-tuned model typically has a significantly higher number of \texttt{(Passed, Passed)} instances compared to the base model. However, for samples that pass the base problems but fail in HumanEval Pro and MBPP Pro, i.e., \texttt{(Failed, Passed)}, the instruct model does not demonstrate notable improvement.
This observation  underscores our argument in \cref{sec:comp}: current instruction-based fine-tuning approaches are insufficiently effective for more complex self-invoking code generation tasks.

\subsection{Chain-of-Thought Prompting}

\begin{table}[t]
\centering
\resizebox{\linewidth}{!}{%
\small

\begin{tabular}{l|c|cc}
\toprule
\textbf{Model}                                       & \textbf{CoT}                 & \textbf{HE Pro} & \textbf{MBPP Pro} \\
\midrule
\multirow{2}{*}{GPT-4o}                     & \no  & 75.0          & 70.9     \\
                                            & \yes & 78.0          & 70.9     \\
\midrule
\multirow{2}{*}{DeepseekV2.5}               & \no  & 73.8          & 71.2     \\
                                            & \yes & 74.4          & 71.4    \\
\midrule
\multirow{2}{*}{Qwen2.5-Coder-32B-ins} & \no  & 70.1          & 69.8     \\
                                            & \yes & 72.0          & 70.1     \\
\midrule
\multirow{2}{*}{Qwen2.5-Coder-7B-ins}  & \no  & 65.9          & 64.8     \\
                                            & \yes & 71.3          & 64.8     \\

\bottomrule
\end{tabular}
}
\caption{The Result with and without CoT on self-invoking code generation benchmarks.}
\label{tab:cot}
\end{table}

\begin{figure}[!t]
    \centering
    \includegraphics[width=0.84\linewidth]{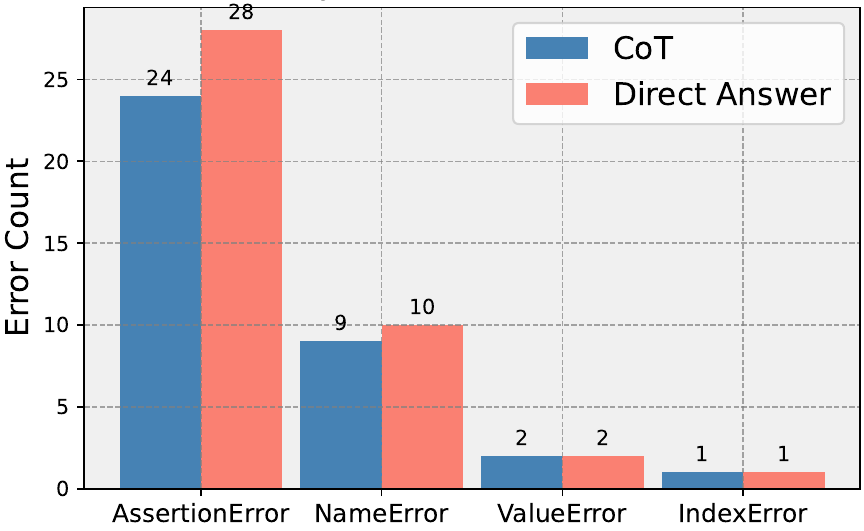}
    \caption{Error types of GPT-4o with and without CoT reasoning on HumanEval Pro.}
    \label{fig:cot-error}
\end{figure}

To evaluate the impact of the model's reasoning ability, we evaluated the performance of GPT-4o, DeepseekV2.5, Qwen2.5-Coder-instruct (7B and 32B) with and without Chain-of-Thought (CoT) prompting~\cite{wei2022chain}  on HumanEval Pro and MBPP Pro. The full prompt we use is shown in \cref{app:prompt}. For CoT prompting, we used the greedy decoding strategy for generation to align the results before. 
As shown in \cref{tab:cot}, 
after applying CoT, the pass@1 of the selected models on HumanEval Pro witnesses a significant improvement. Notably, the accuracy of GPT-4o increases from 75.0\% to 78.0\%. On MBPP Pro, although the model does not show a significant improvement, it still maintains its original performance level, indicating that CoT can enhance the accuracy of model-generated code to a notable degree.

\textbf{CoT could help Code LLMs to generate more reliable code when scheduling across multiple code-related problems.} 
To further study which aspects of code LLM can be improved by CoT, 
we use Python to run the code generated by GPT4o with and without CoT, and present the number of all error types that occurred in \cref{fig:cot-error}. We have two main observations: (1) With CoT prompting, the \emph{AssertionError} number decreases from 28 to 24. This indicates that CoT prompting enables the model to generate code that more frequently passes test cases. (2) The \emph{NameError} number decreases, which indicates that CoT prompting helps the model produce more self-contained code snippets and reduces the use of undefined variables.
These findings highlight that CoT prompting could help LLMs to generate more accurate and reliable solution on self-invoking code generation task.

\begin{figure}[!t]
    \centering
    \includegraphics[width=\linewidth]{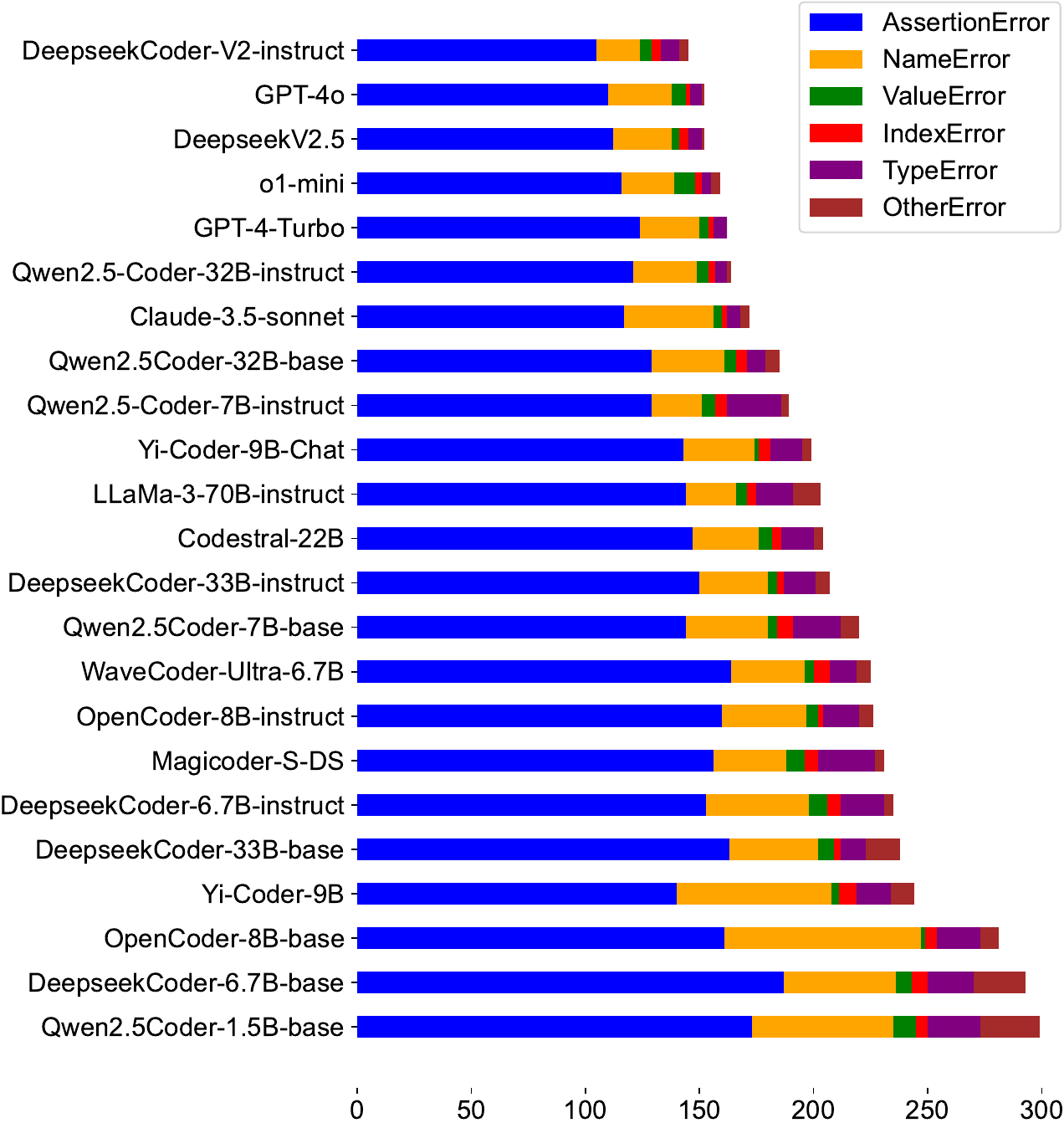}
    \caption{Statistics of error type across different LLMs on HumanEval Pro and MBPP Pro. We sum up all kinds of errors on the two benchmarks. Exact number is shown in \cref{tab:error}.}
    \label{fig:stats}
\end{figure}

\subsection{Error Analysis}
In order to further understand the failure modes across different LLMs, we analyze the errors encountered in code generated by different LLMs for HumanEval Pro and MBPP Pro problems and categorize them by error type. The result is shown in \cref{fig:stats}.
Primarily, \textit{AssertionErrors} constitute the primary source of errors for all models on self-invoking code generation task, which suggests that the majority of errors are still due to failing test cases.
Secondly, the \textit{NameErrors}, which is often caused by the undefined variable or function, contribute significantly to the error rate. This suggests that despite the function infomation being provided in the prompt, many functions still fail to generate the correct function header. This may indicate that the LLM has issues with understanding or correctly utilizing the provided information. Finally, we also found that some \textit{TypeErrors} and \textit{ValueErrors} accounted for a relatively small proportion of errors, which shows that LLM still has some deficiencies in handling variable types and usage when generating self-invoking code.

\section{Generalization Study of Self-invoking Code Generation}

\subsection{BigCodeBench-Lite Pro Benchmark}
To study self-invoking code generation on a wider range of programming problems,
we construct BigCodeBench-Lite Pro, a small self-invoking code generation benchmark derived from BigCodeBench~\cite{zhuo2024bigcodebench}. 
We first construct the BigCodeBench-Lite benchmark by selecting 57 problems with solve rate between 50\% and 70\% from BigCodeBench\footnote{We use reported statistics in \url{https://huggingface.co/datasets/bigcode/bigcodebench-solve-rate}.}. For each examples in BigCodeBench-Lite, we then curate the corresponding self-invoking problem as well as test cases, following the same procedure described in \cref{sec3:bench}. 
After further filtering by human experts, BigCodeBench-Lite Pro contains 57 self-invoking programming problems from different topics.

\begin{table}[t!]
\small
\centering
\begin{tabular}{l|cc}
\toprule
\textbf{Model}              & \textbf{BCB-Lite}         & \textbf{Pro (\%)} \\
\midrule
GPT-4o                      & 64.9 & 52.6                  \\
GPT4-Turbo                  & 61.4 & 52.6                  \\
Claude-3.5-sonnet           & 73.7 & 50.9                  \\
DeepseekV2.5                & 80.7 & 50.9                  \\
\midrule
Qwen2.5Coder-1.5B-base      & 50.9 & 15.8                  \\
Qwen2.5Coder-1.5B-instruct  & 50.9 & 10.5                  \\
\midrule
OpenCoder-8B-base           & 56.1 & 10.5                  \\
OpenCoder-8B-instruct       & 75.4 & 22.8                  \\
\midrule
DeepseekCoder-6.7B-base     & 59.6 & 35.1                  \\
DeepseekCoder-6.7B-instruct & 56.1 & 35.1                  \\
WaveCoder-Ultra-6.7B        & 61.4 & 26.3                  \\
Magicoder-S-DS-6.7B         & 50.9 & 33.3                  \\
\midrule
Yi-Coder-9B                 & 57.9 & 21.1                  \\
Yi-Coder-9B-Chat            & 66.7& 31.6                  \\
\midrule
Qwen2.5Coder-7B-base        & 59.6 & 38.6                  \\
Qwen2.5Coder-7B-instruct    & 64.9 & 35.1             
\\
\midrule
DeepseekCoder-33B-base      & 71.9 & 38.6                  \\
DeepseekCoder-33B-instruct  & 80.7 & 43.9                  \\
\midrule
Qwen2.5Coder-32B-base       & 68.4 & 49.1                  \\
Qwen2.5Coder-32B-instruct   & 80.7 & 52.6                  \\
\midrule
Codestral-22B               & 78.9 & 54.4                  \\
\midrule
QwQ-32B-preview             & 86.0 & 59.6                  \\
\bottomrule
\end{tabular}
\caption{Passing rate (\%) of LLMs on BigCodeBench (BCB)-Lite and BCB-Lite-Pro. A dataset example of BCB-Lite-Pro is shown in \cref{app:bcb-exp}.}
\label{tab:bcb}
\end{table}

\subsection{Results Analysis}
We evaluate a set of LLMs on BigCodeBench-Lite Pro. \cref{tab:bcb} presents the results (pass@1) of various Proprietary and Open-source LLMs,  highlighting the following observations:
(1) Although the base problems we selected has a solving rate of between 50\% and 70\% on BigCodeBench, only a small number of models in \cref{tab:bcb} have a passing rate of more than 50\% on BigCodeBench-Lite Pro. This highlights the difficulty of the self-invoking code generation task.
(2) The instruction-tuned models still demonstrate marginal improvements (sometimes decrease) compared to base models, which also reinforces our argument in \cref{sec:comp}.
\section{Conclusion}
We present HumanEval Pro, MBPP Pro as well as BigCodeBench-Lite Pro, a series of benchmarks to evaluate LLMs on self-invoking code generation task where the LLMs are employed to solve the base problem and use its solution to address more complex problems. Through extensive evaluation of over 20 LLMs, we found that while these models have made significant progress in traditional code generation tasks, they still struggle with more complex self-invoking code generation tasks. 
Furthermore, we provide extensive comparison and analysis between existing instruct model and base model. 
HumanEval Pro and MBPP Pro are positioned to serve as valuable benchmarks for code-related evaluations and to inspire future LLM development by shedding light on current model shortcomings and encouraging innovation in training methodologies.

\section*{Limitations}
In this paper, we present HumanEval Pro and MBPP Pro, a series of benchmarks evaluate LLMs on self-invoking code generation task. One limitation is that the programming languages of our benchmarks 
only includes Python due to the intrinsic limitation of original HumanEval and MBPP. Secondly, although the models have shown shortcomings in the self-invoking problem, the diversity of existing self-invoking problems in HumanEval Pro and MBPP Pro is still subject to the constraints of the original problems. Hence, future work should pay more attention to more diverse and multi-lingual self-invoking problem benchmarks.

\bibliography{acl}

\appendix
\onecolumn

\addtocontents{toc}{\protect\setcounter{tocdepth}{3}}

\renewcommand{\contentsname}{\Large Appendix Contents}
\hypersetup{linkcolor=black}
\tableofcontents
\clearpage

\input{appendix-random-results}
\input{appendix-link}

\input{appendix-self-invoking-example}
\input{appendix-discussion-solutions}
\input{appendix-prompts}

\input{appendix-error-analysis}

\end{document}

%% file: appendix-random-results.tex

\section{Detailed Results}
\label{sec:res}

\begin{table}[h]
\centering

\begin{tabular}{l|cc}
\toprule
\textbf{Model}                       & \textbf{HumanEval Pro} (0-shot) & \textbf{MBPP Pro} (0-shot) \\
\midrule
LLaMA-3.1-8B-base           & 25.0                   & 36.5              \\
LLaMA-3.1-8B-instruct       & 45.7                   & 53.7              \\
\midrule
LLaMA-3.1-70B-base          & 40.9                   & 57.4              \\
LLaMA-3.1-70B-instruct      & 60.4                   & 63.8              \\
\midrule
Qwen-2.5-72B-base           & 62.2                   & 65.3              \\
Qwen-2.5-72B-instruct       & 68.9                   & 68.8              \\
\midrule
QwQ-32B-preview             & 72.0                   & 67.5               \\
LLaMA-3.3-70B-instruct      & 67.1                   & 64.6              \\
Mistral-Large-instruct-2411 & 75.0                   & 69.3              \\
\bottomrule
\end{tabular}
\caption{Results of Other LLMs on HumanEval Pro and MBPP Pro (greedy decoding).}
\label{tab:other-res}
\end{table}

\begin{table}[h]
\centering

\begin{tabular}{l|ccc|ccc}
\toprule
\multirow{2}{*}{\textbf{Model}}    & \multicolumn{3}{c|}{\textbf{HumanEval Pro}} & \multicolumn{3}{c}{\textbf{MBPP Pro}} \\
                          & \textbf{pass@1}    & \textbf{pass@5}    & \textbf{pass@10}   & \textbf{pass@1}  & \textbf{pass@5}  & \textbf{pass@10}  \\

\midrule
DeepseekCoder-6.7B-base & 38.0      & 50.9      & 54.7      & 51.6    & 60.4    & 63.1    \\
DeepseekCoder-6.7B-instruct & 55.9      & 64.1      & 66.5      & 55.2    & 62.6    & 64.9    \\
Magicoder-S-DS-6.7B & 55.1      & 62.7      & 65.1      & 57.7    & 64.9    & 67.2    \\
WaveCoder-Ultra-6.7B & 55.7      & 61.4      & 63.0      & 58.2    & 64.4    & 66.3    \\
\midrule
DeepseekCoder-33B-base & 49.4      & 60.8      & 65.2      & 59.1    & 67.2    & 69.3    \\
DeepseekCoder-33B-instruct & 59.1      & 68.6      & 71.3      & 63.4    & 70.6    & 72.9    \\
\midrule
Qwen2.5-Coder-7B-base & 51.8      & 62.1      & 66.2      & 61.3    & 69.9    & 72.3    \\
Qwen2.5-Coder-7B-instruct & 65.7      & 72.5      & 75.0      & 64.2    & 70.5    & 72.6    \\
\midrule
OpenCoder-9B-base & 44.5      & 56.2      & 59.9      & 54.8    & 62.9    & 65.0    \\
OpenCoder-9B-instruct & 59.8      & 68.5      & 70.8      & 58.1    & 63.7    & 65.1    \\
\midrule
Yi-Coder-9B-base & 47.9      & 59.0      & 61.9      & 59.6    & 67.7    & 69.7    \\
Yi-Coder-9B-chat & 59.7      & 66.4      & 67.9      & 65.0    & 69.8    & 71.2    \\
\midrule
Codestral-22B & 59.5      & 66.2      & 67.7      & 63.2    & 67.7    & 68.9    \\
\midrule
Qwen2.5-Coder-32B-base & 62.4      & 70.3      & 72.2      & 67.6    & 75.0    & 76.9    \\
Qwen2.5-Coder-32B-instruct & 69.2      & 72.3      & 73.3      & 70.6   & 74.7    & 76.0    \\
QwQ-32B-preview & 70.9 & 77.7 & 79.5 & 67.0 & 73.0 & 74.5 \\
\bottomrule

\end{tabular}
\caption{The results of different models on HumanEval Pro and MBPP Pro . We generate 20 samples for each problems with random sampling strategy where temperature is set to 0.2 and top\_p is set to 0.95.}
\end{table}

\section{Example in Benchmark Construction}

\begin{figure}[h]
    \centering
    \includegraphics[width=0.9\linewidth]{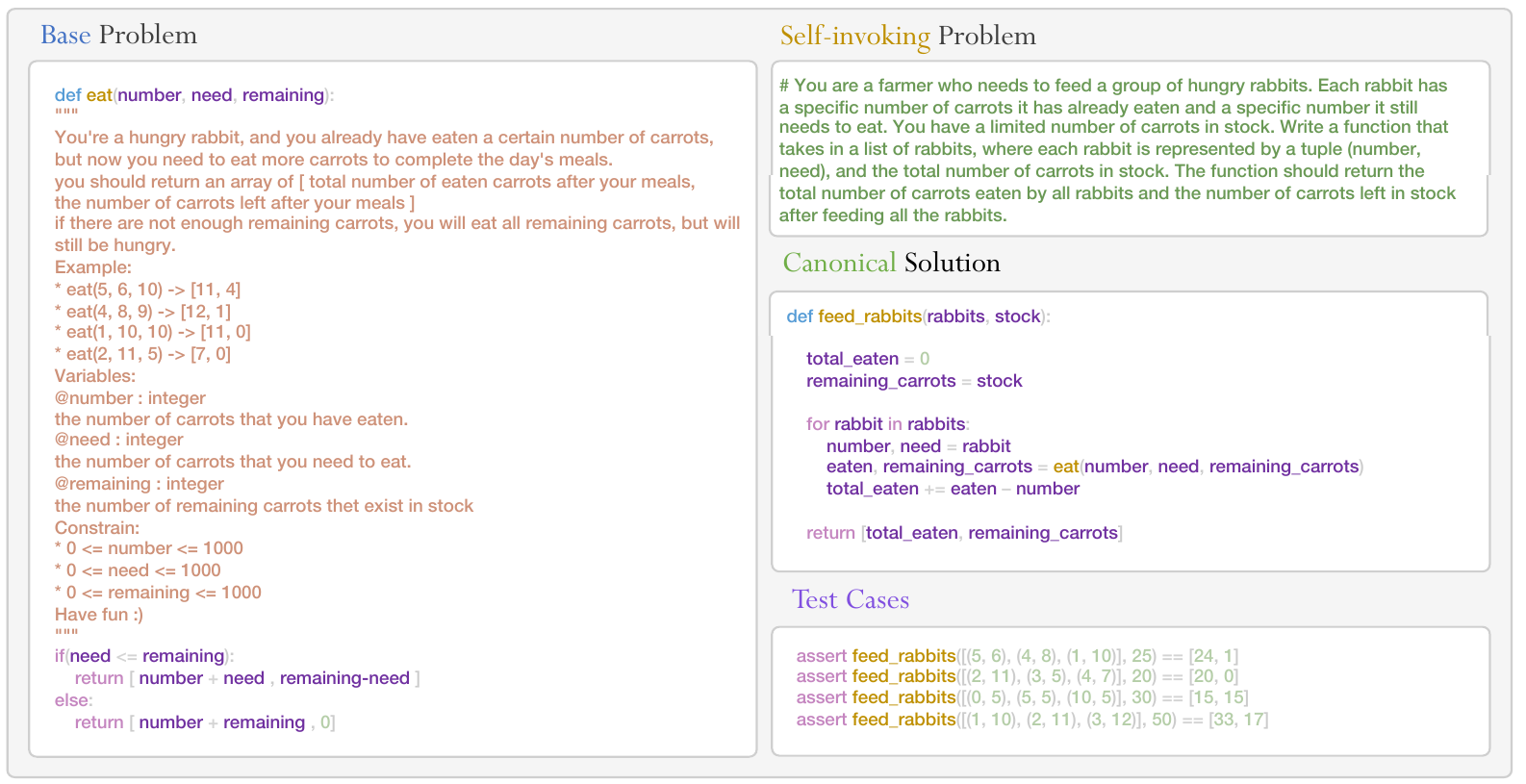}
    \caption{An example of self-invoking problems in HumanEval
Pro }
    \label{fig:bc_exp}
\end{figure}

%% file: appendix-link.tex
\newpage
\section{Model Information}
\label{app:model}
\begin{table}[h!]
    \centering
    \small
    \begin{tabular}{ll}
\toprule
\textbf{Model Name} & \textbf{API Name} \\
\midrule
    O1-mini & \texttt{o1-mini-2024-09-12} \\
    GPT-4o & \texttt{gpt-4o-2024-08-06} \\
    GPT-4-Turbo & \texttt{gpt-4-turbo-2024-04-09} \\
    Claude-3.5-sonnet & \texttt{claude-3-5-sonnet-20241022} \\
    Deepseek-V2.5 & \texttt{deepseek-chat}\\
\bottomrule
\end{tabular}

\vspace{0.3cm}

\begin{tabular}{ll}
\toprule
\textbf{Model Name} & \textbf{HuggingFace URL} \\
\midrule

DeepseekCoder-V2-instruct & \url{https://huggingface.co/deepseek-ai/DeepSeek-Coder-V2-Instruct} \\
Qwen2.5-Coder-1.5B-base & \url{https://huggingface.co/Qwen/Qwen2.5-Coder-1.5B} \\
Qwen2.5-Coder-1.5B-instruct & \url{https://huggingface.co/Qwen/Qwen2.5-Coder-1.5B-Instruct} \\
DeepseekCoder-6.7B-base & \url{https://huggingface.co/deepseek-ai/deepseek-coder-6.7b-base} \\
DeepseekCoder-6.7B-instruct & \url{https://huggingface.co/deepseek-ai/deepseek-coder-6.7b-instruct} \\
Magicoder-S-DS-6.7B & \url{https://huggingface.co/ise-uiuc/Magicoder-S-DS-6.7B} \\
WaveCoder-Ultra-6.7B & \url{https://huggingface.co/microsoft/wavecoder-ultra-6.7b} \\
Qwen2.5-Coder-7B-base & \url{https://huggingface.co/Qwen/Qwen2.5-Coder-7B} \\
Qwen2.5-Coder-7B-instruct & \url{https://huggingface.co/Qwen/Qwen2.5-Coder-7B-Instruct} \\
OpenCoder-8B-base & \url{https://huggingface.co/infly/OpenCoder-8B-Base} \\
OpenCoder-8B-instruct & \url{https://huggingface.co/infly/OpenCoder-8B-Instruct} \\
Yi-Coder-9B-base & \url{https://huggingface.co/01-ai/Yi-Coder-9B} \\
Yi-Coder-9B-chat & \url{https://huggingface.co/01-ai/Yi-Coder-9B-Chat} \\
Codestral-22B-v0.1 & \url{https://huggingface.co/mistralai/Codestral-22B-v0.1} \\
DeepseekCoder-33B-base & \url{https://huggingface.co/deepseek-ai/deepseek-coder-33b-base} \\
DeepseekCoder-33B-instruct & \url{https://huggingface.co/deepseek-ai/deepseek-coder-33b-instruct} \\
Qwen2.5-Coder-32B-base & \url{https://huggingface.co/Qwen/Qwen2.5-Coder-32B} \\
Qwen2.5-Coder-32B-instruct & \url{https://huggingface.co/Qwen/Qwen2.5-Coder-32B-Instruct} \\
LLaMA3-70B-instruct & \url{https://huggingface.co/meta-llama/Meta-Llama-3-70B-Instruct} \\
QwQ-32B-Preview & \url{https://huggingface.co/Qwen/QwQ-32B-Preview} \\
LLaMA3.1-8B-base & \url{https://huggingface.co/meta-llama/Llama-3.1-8B} \\
LLaMA3.1-8B-instruct & \url{https://huggingface.co/meta-llama/Llama-3.1-8B-Instruct} \\
LLaMA3.1-70B-base & \url{https://huggingface.co/meta-llama/Llama-3.1-70B} \\
LLaMA3.1-70B-instruct & \url{https://huggingface.co/meta-llama/Llama-3.1-70B-Instruct} \\
Qwen2.5-72B-base & \url{https://huggingface.co/Qwen/Qwen2.5-72B} \\
Qwen2.5-72B-instruct & \url{https://huggingface.co/Qwen/Qwen2.5-72B-Instruct} \\
\bottomrule
\end{tabular}

\caption{The corresponding  API names and  HuggingFace model URLs for the evaluated  models are listed in \cref{tab:result}.
}

\end{table}

%% file: appendix-self-invoking-example.tex
\clearpage
\section{Comparison between HumanEval (Pro), MBPP (Pro) and BigCodeBench-Lite (Pro)}
\begin{figure}[h!]
    \centering
    \includegraphics[width=0.9\linewidth]{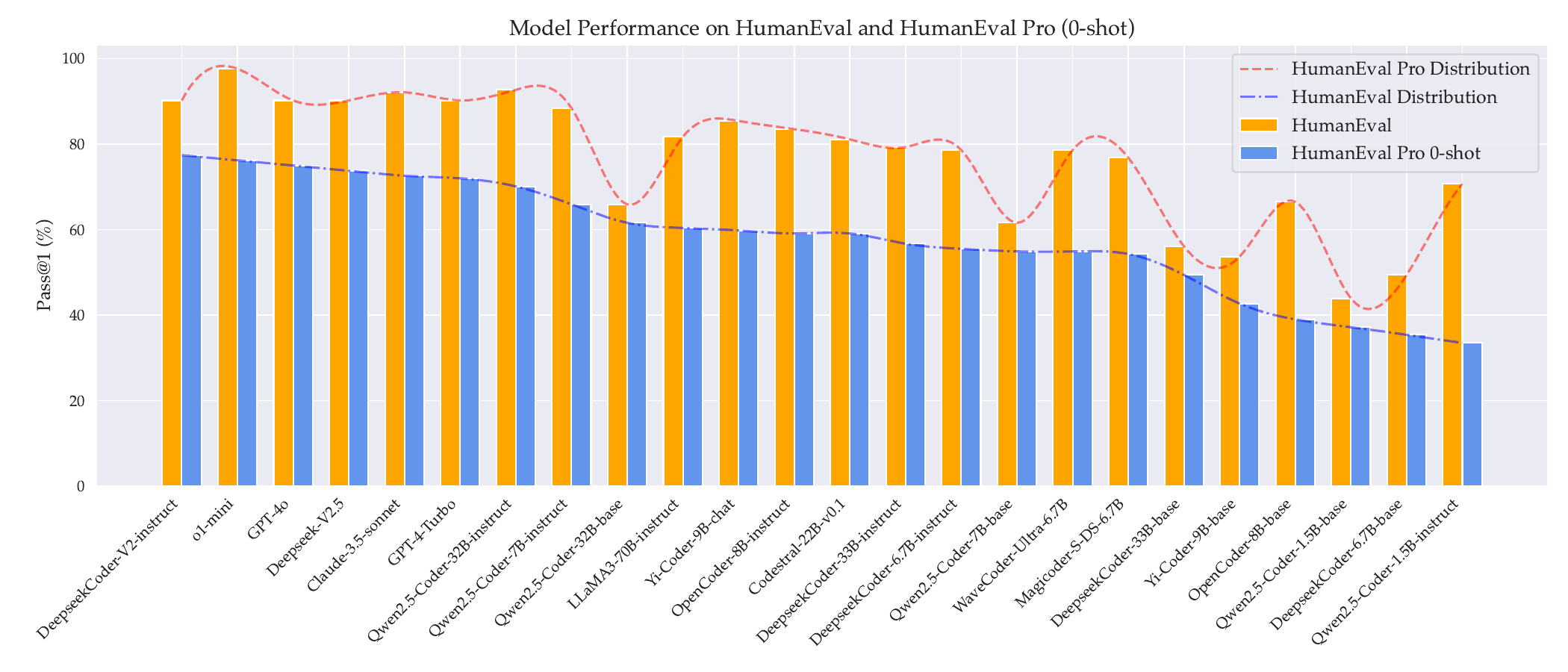}
    \includegraphics[width=0.9\linewidth]{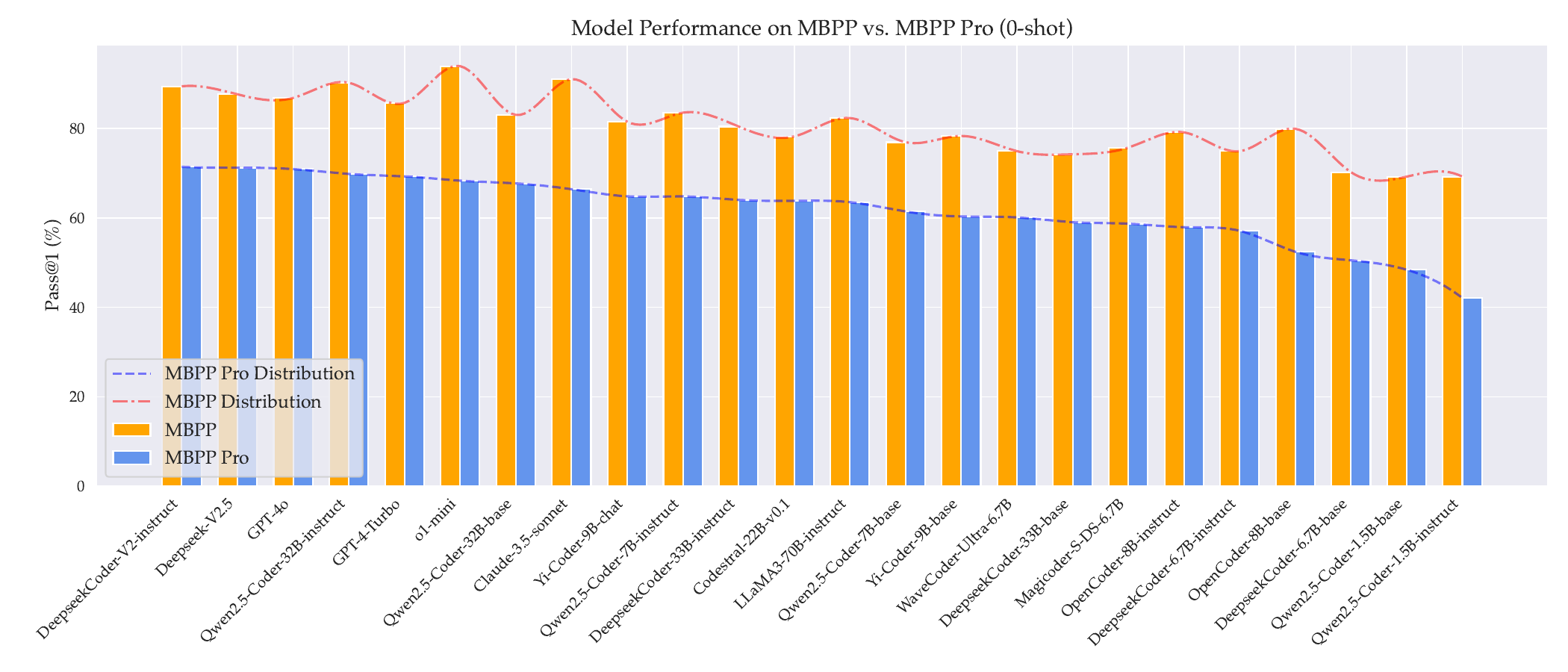}
    \includegraphics[width=0.9\linewidth]{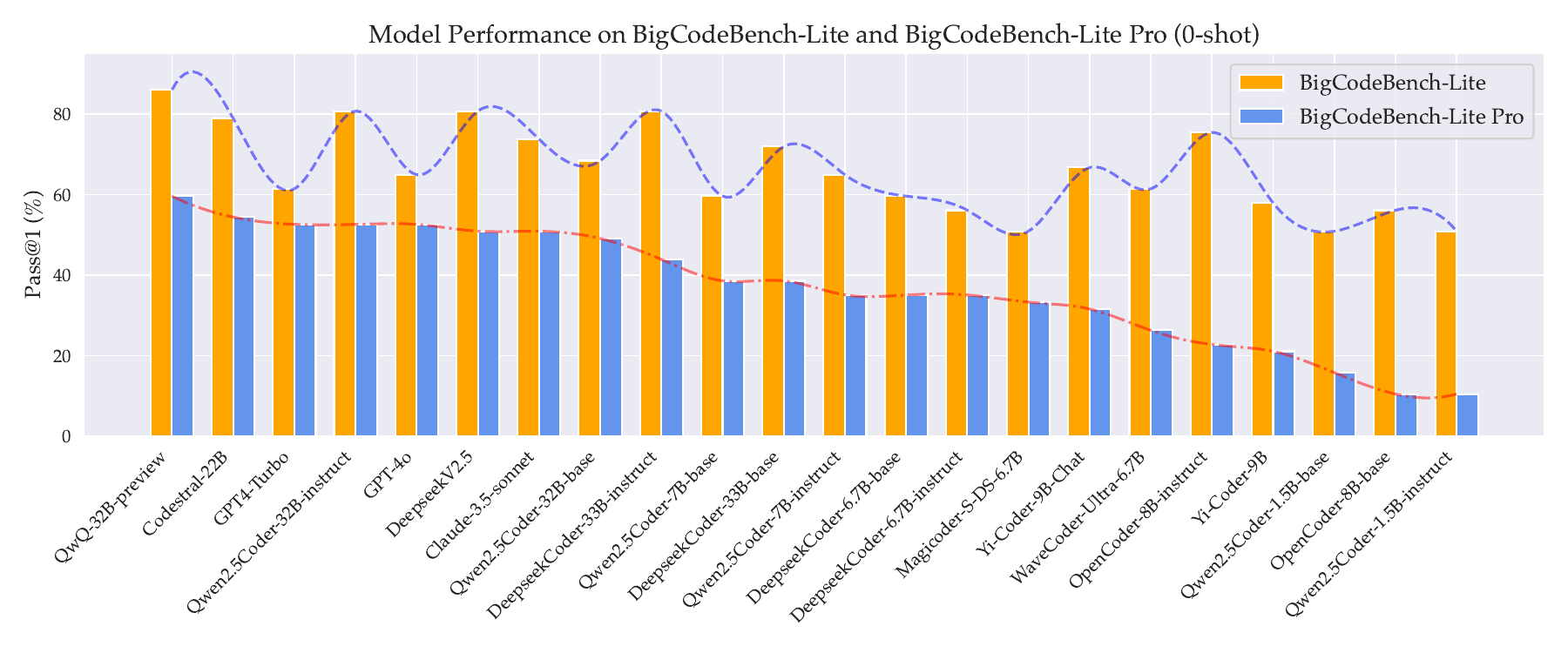}

    \caption{Comparison between HumanEval Family, MBPP Family and BigCodeBench-Lite Family.}
    \label{fig:app-bench-comp}
\end{figure}

%% file: appendix-discussion-solutions.tex
\newpage
\section{Discussion about Self-invoking Problems and Solutions}
We analyze the complexity comparison between a base problem and its self-invoking counterpart by examining the line count of their canonical solutions. The line count serves as a proxy for the complexity of each problem. By comparing the number of lines required to solve the base problem with those needed for the self-invoking version, we gain insight into how the introduction of self-invocation affects the overall complexity. Generally, self-invoking problems, which often involve recursion or similar constructs, may require more lines of code to handle additional logic and edge cases, thereby increasing the complexity. This comparison helps in understanding the additional computational and conceptual challenges introduced by self-invocation.
\begin{figure}[h]
    \centering
    \includegraphics[width=0.85\linewidth]{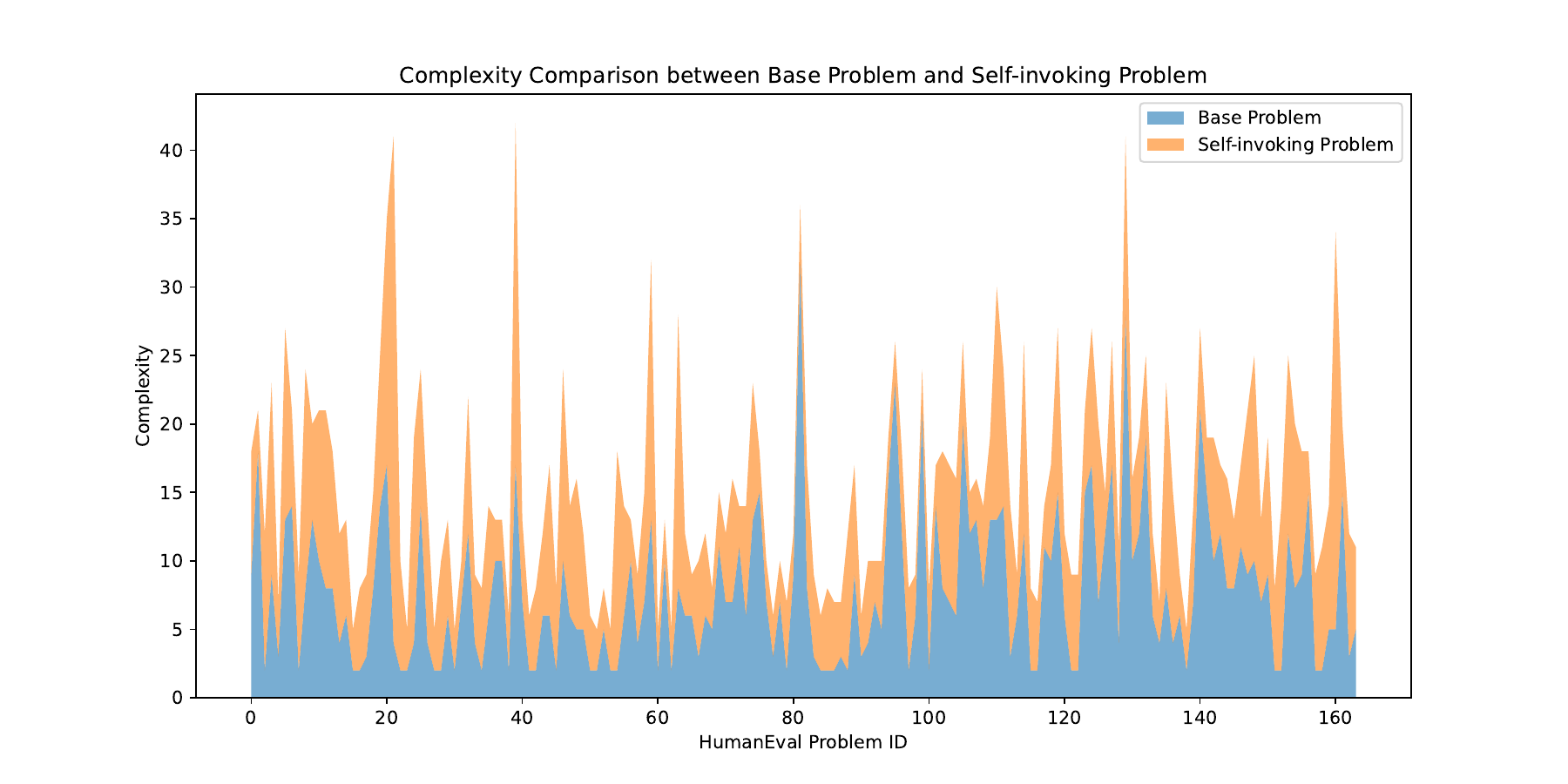}
    \includegraphics[width=0.85\linewidth]{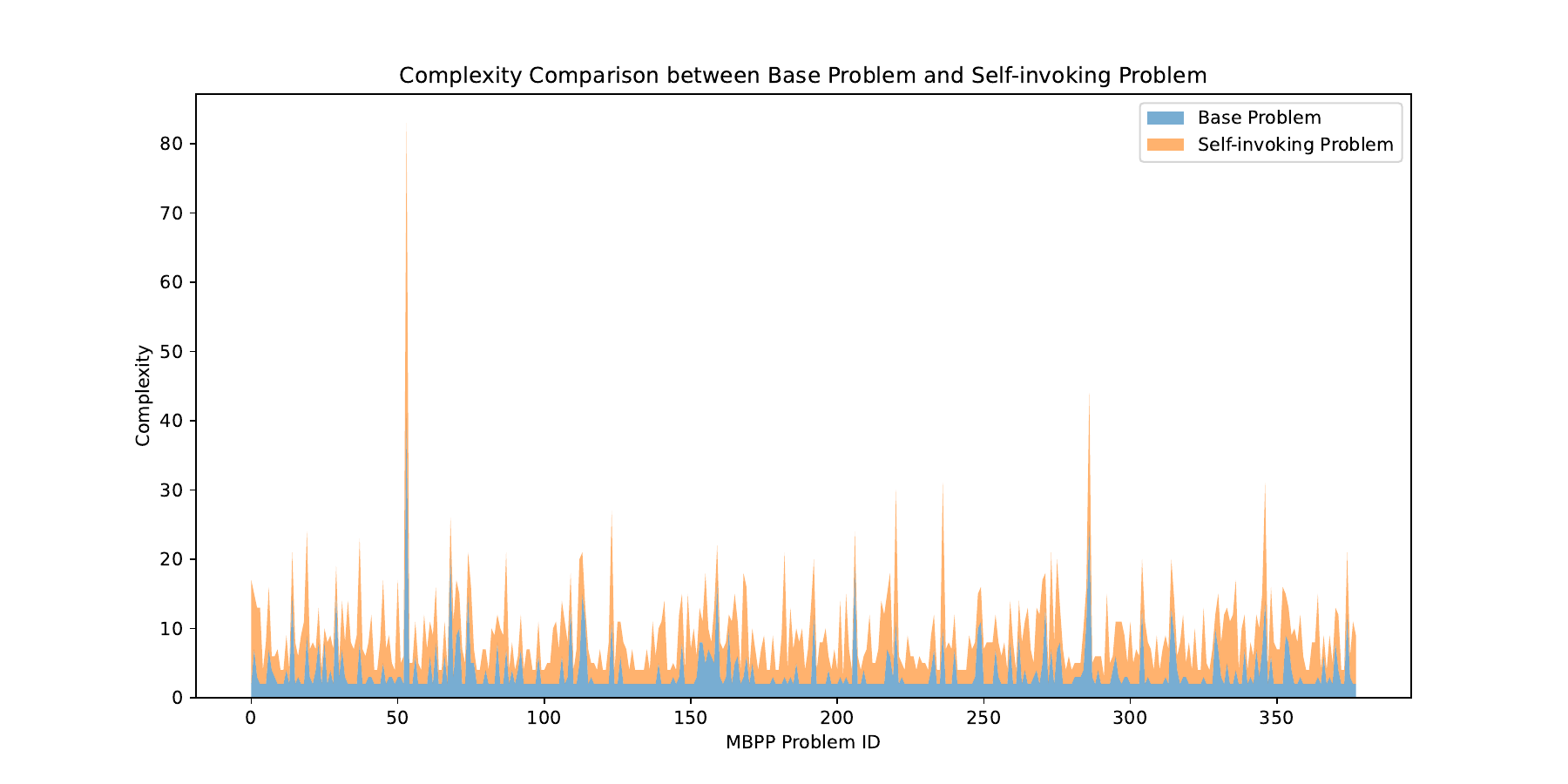}
    \caption{Complexity comparison between base problem and self-invoking problem. We use the line count of the canonical solution for both the base problem and the self-invoking problem as a measure of the problem's complexity.}
    \label{fig:dis-sol}
\end{figure}

%% file: appendix-prompts.tex
\newpage
\section{Prompts}
\subsection{Prompts for Benchmark Construction}
\label{app:bench-prompt}
We set the prompt in our benchmark construction as follows:
\definecolor{mycolor}{RGB}{208,223,230}
\begin{tcolorbox}[colback=mycolor]
    \textbf{Prompt for benchmark construction:} 
    
    I'll give you a raw programming question and its solution, please generate a new problem that requires multiple calls to the raw question  to solve, and generate the solution in new\_solution.
    
    Please return with json format including 3 keys:     'new\_problem','new\_solution', 'test\_input\', 
    I'll use json.loads() to transform it to dict type.
    
    To solve new\_problems, new\_solution should include the multiple function calls of raw question. So new\_problems will be not only a related problem but also a more complex problem than raw problem.
    
    raw problem:
    
    \{raw problem\}
    
    raw solution:
    
    \{raw solution\}
\end{tcolorbox}

\subsection{Prompts for Evaluation}
\label{app:prompt}
We set the prompt in our evaluation as follows:
\definecolor{mycolor}{RGB}{208,223,230}
\begin{tcolorbox}[colback=mycolor]
    \textbf{Prompt of 0-shot:} 
        You are an exceptionally intelligent coding assistant that consistently delivers accurate and reliable responses to user instructions. Write a solution of python file to the following problems, the solution of the
        second problem requires single or multiple calls to the first 
        
        @@ Instruction
        
        \{base problem\}
        
        \{self-invoking problem\}
        
        @@ Response
    \\[0.3cm]
    \textbf{Prompt of 1-shot:} 
        You are an exceptionally intelligent coding assistant that consistently delivers accurate and reliable responses to user instructions. Write a solution of python file to the following problems, the solution of the
        second problem requires single or multiple calls to the first solution
        
        @@ Instruction
        
        \{base problem\}
        
        \{self-invoking problem\}

        \{example\}
        
        @@ Response
    \\[0.3cm]
     \textbf{Prompt with Chain-of-Thought:}  
    
        You are an exceptionally intelligent coding assistant that consistently delivers accurate and reliable responses to user instructions. Write a solution of python file to the following problems, the solution of the
        second problem requires single or multiple calls to the first 
        
        @@ Instruction
        
        \{base problem\}
        
        \{self-invoking problem\}
        
        @@ Response
        
        Let's think step by step.
\end{tcolorbox}

%% file: appendix-error-analysis.tex
\section{Examples of Different Error Types}
\subsection{Examples of AssertionError}
\label{app:assertionerror}
\begin{lstlisting}[caption=AssertionError in HumanEval/24-generated by Qwen2.5Coder-7B-base,basicstyle=\ttfamily]
Status: AssertionError
Traceback: assert largest_common_divisor([15, 30, 45]) == 15
Explanation: expected 15 but got 5.
\end{lstlisting}
\begin{lstlisting}[basicstyle=\ttfamily]
def largest_divisor(n: int) -> int:
    """ For a given number n, find the largest number that divides n evenly, smaller than n
    >>> largest_divisor(15)
    5
    """
    for i in range(n-1, 0, -1):
        if n % i == 0:
            return i
    return 1

def largest_common_divisor(numbers: list) -> int:
    """ For a given list of integers, find the largest number that divides each of the numbers in the list evenly, smaller than the smallest number in the list
    >>> largest_common_divisor([15, 30, 45])
    15
    """
    smallest = min(numbers)
    for i in range(smallest-1, 0, -1):
        if all(n % i == 0 for n in numbers):
            return i
    return 1

assert largest_common_divisor([15, 30, 45]) == 15
assert largest_common_divisor([15, 30, 45]) == 15
assert largest_common_divisor([10, 20, 30]) == 10
assert largest_common_divisor([7, 14, 21]) == 7
assert largest_common_divisor([16, 32, 64]) == 16
assert largest_common_divisor([9, 18, 27]) == 9
\end{lstlisting}

\subsection{Examples of NameError}

\label{app:nameerror}
\begin{lstlisting}[caption=NameError in HumanEval/14-generated by Qwen2.5Coder-7B-instruct,basicstyle=\ttfamily]
Status: NameError
Traceback: assert all_prefixes_for_list(['abc', 'def']) == [['a', 'ab', 'abc'], ['d', 'de', 'def']]
Explanation: The expected function name is 'all_prefixes_for_list', but 'all_prefixes_of_strings' is generated though this information is given in prompt.
\end{lstlisting}
\begin{lstlisting}[basicstyle=\ttfamily]
from typing import List
def all_prefixes(string: str) -> List[str]:
    """ Return list of all prefixes from shortest to longest of the input string
    >>> all_prefixes('abc')
    ['a', 'ab', 'abc']
    """
    return [string[:i] for i in range(1, len(string) + 1)]

def all_prefixes_of_strings(strings: List[str]) -> List[List[str]]:
    """ Return list of lists where each sublist contains all prefixes of the corresponding string in the input list, sorted from shortest to longest. If the input list is empty, return an empty list.
    >>> all_prefixes_of_strings(['abc', 'def', 'ghi'])
    [['a', 'ab', 'abc'], ['d', 'de', 'def'], ['g', 'gh', 'ghi']]
    """
    return [all_prefixes(s) for s in strings]

assert all_prefixes_for_list(['abc', 'def']) == [['a', 'ab', 'abc'], ['d', 'de', 'def']]
assert all_prefixes_for_list(['']) == [[]]
assert all_prefixes_for_list([]) == []
assert all_prefixes_for_list(['single']) == [['s', 'si', 'sin', 'sing', 'singl', 'single']]
\end{lstlisting}

\subsection{Examples of ValueError}

\label{app:valueerror}
\begin{lstlisting}[caption=ValueError in HumanEval/153-generated by O1-mini,basicstyle=\ttfamily]
Status: ValueError
Traceback: 
>>> assert Strongest_Extensions({'Slices': ['SErviNGSliCes', 'Cheese', 'StuFfed'], 'my_class': ['AA', 'Be', 'CC']}) == ['Slices.SErviNGSliCes', 'my_class.AA']
>>> for class_name, extensions in classes_with_extensions
Explanation: ValueError: too many values to unpack (expected 2)
\end{lstlisting}
\begin{lstlisting}[basicstyle=\ttfamily]

def Strongest_Extension(class_name, extensions):
    """
    Determines the strongest extension for a given class based on the strength calculation.
    
    The strength of an extension is calculated as the number of uppercase letters 
    minus the number of lowercase letters in the extension's name.
    
    If multiple extensions have the same strength, the first one in the list is chosen.
    
    Parameters:
    - class_name (str): The name of the class.
    - extensions (List[str]): A list of extension names.
    
    Returns:
    - str: A string in the format "ClassName.StrongestExtensionName".
    
    Example:
    >>> Strongest_Extension('my_class', ['AA', 'Be', 'CC'])
    'my_class.AA'
    """
    if not extensions:
        return f"{class_name}.None"
    
    def calculate_strength(extension):
        cap = sum(1 for char in extension if char.isupper())
        sm = sum(1 for char in extension if char.islower())
        return cap - sm
    
    strongest_extension = extensions[0]
    max_strength = calculate_strength(strongest_extension)
    
    for ext in extensions[1:]:
        strength = calculate_strength(ext)
        if strength > max_strength:
            strongest_extension = ext
            max_strength = strength
    return f"{class_name}.{strongest_extension}"

def Strongest_Extensions(classes_with_extensions):
    """
    Determines the strongest extensions for a list of classes.
    
    For each class, it finds the strongest extension based on the strength calculation.
    If a class has no extensions, it is represented as "ClassName.None".
    
    Parameters:
    - classes_with_extensions (List[Tuple[str, List[str]]]): 
      A list where each element is a tuple containing a class name and a list of its extensions.
    
    Returns:
    - List[str]: A list of strings in the format "ClassName.StrongestExtensionName".
    
    Example:
    >>> Strongest_Extensions([
    ...     ('my_class', ['AA', 'Be', 'CC']),
    ...     ('Slices', ['SErviNGSliCes', 'Cheese', 'StuFfed']),
    ...     ('EmptyClass', [])
    ... ])
    ['my_class.AA', 'Slices.SErviNGSliCes', 'EmptyClass.None']
    """
    result = []
    for class_name, extensions in classes_with_extensions:
        if extensions:
            strongest = Strongest_Extension(class_name, extensions)
        else:
            strongest = f"{class_name}.None"
        result.append(strongest)
    return result

assert Strongest_Extensions({'Slices': ['SErviNGSliCes', 'Cheese', 'StuFfed'], 'my_class': ['AA', 'Be', 'CC']}) == ['Slices.SErviNGSliCes', 'my_class.AA']
assert Strongest_Extensions({'Empty': [], 'Single': ['A']}) == ['Empty.None', 'Single.A']
assert Strongest_Extensions({'Mixed': ['aA', 'Bb', 'Cc'], 'AllCaps': ['AA', 'BB', 'CC']}) == ['Mixed.aA', 'AllCaps.AA']
assert Strongest_Extensions({'NoExtensions': []}) == ['NoExtensions.None']
assert Strongest_Extensions({}) == []  # Check for empty input

\end{lstlisting}

\subsection{Examples of IndexError}

\label{app:indexerror}
\begin{lstlisting}[caption=IndexError in HumanEval/73-generated by GPT-4o,basicstyle=\ttfamily]
Status: IndexError
Traceback: 
>>> assert smallest_change_2d([[1,2,3,5,4,7,9,6], [1,2,3,4,3,2,2], [1,2,3,2,1]]) == 7
>>> column = [arr[row][col] for row in range(rows)]
Explanation: IndexError: list index out of range
\end{lstlisting}
\begin{lstlisting}[basicstyle=\ttfamily]

def smallest_change(arr):
    """
    Given an array arr of integers, find the minimum number of elements that
    need to be changed to make the array palindromic. A palindromic array is an array that
    is read the same backwards and forwards. In one change, you can change one element to any other element.
    """
    n = len(arr)
    changes = 0
    for i in range(n // 2):
        if arr[i] != arr[n - i - 1]:
            changes += 1
    return changes

def smallest_change_2d(arr):
    """
    Given a 2D array of integers, find the minimum number of elements that need to be changed to make each row
    and each column of the array palindromic. A palindromic row or column is a row or column that is read the same
    backwards and forwards. In one change, you can change one element to any other element.
    """
    if not arr or not arr[0]:
        return 0

    rows = len(arr)
    cols = len(arr[0])

    # Calculate changes needed for each row
    row_changes = sum(smallest_change(row) for row in arr)

    # Calculate changes needed for each column
    col_changes = 0
    for col in range(cols):
        column = [arr[row][col] for row in range(rows)]
        col_changes += smallest_change(column)

    return row_changes + col_changes

assert smallest_change_2d([[1,2,3,5,4,7,9,6], [1,2,3,4,3,2,2], [1,2,3,2,1]]) == 7
assert smallest_change_2d([[1,2,3,4], [4,3,2,1], [1,2,2,1], [1,2,3,4]]) == 8
assert smallest_change_2d([[1,2,3,2,1], [1,2,3,2,1], [1,2,3,2,1], [1,2,3,2,1]]) == 0
\end{lstlisting}

\subsection{Examples of TypeError}

\label{app:typeerror}
\begin{lstlisting}[caption=TypeError in MBPP/164-generated by DeepseekCoder-V2-instruct,basicstyle=\ttfamily]
Status: TypeError
Traceback: 
>>> assert total_and_average_volume([{'radius': 3, 'height': 5}, {'radius': 4, 'height': 6}]) == (147.6548547187203, 73.82742735936014)
>>> total_volume += volume_cone(r, h)
>>> return (math.pi * r**2 * h) / 3

Explanation: TypeError: unsupported operand type(s) for ** or pow(): 'str' and 'int'
\end{lstlisting}
\begin{lstlisting}[basicstyle=\ttfamily]
# Write a function to find the volume of a cone.
import math
def volume_cone(r,h):
    return (math.pi * r**2 * h) / 3

# Given a list of cones with their respective radii and heights, write a function to calculate the total volume of all cones and the average volume per cone.

def total_and_average_volume(cones):
    total_volume = 0
    for cone in cones:
        r, h = cone
        total_volume += volume_cone(r, h)
    average_volume = total_volume / len(cones)
    return total_volume, average_volume

assert total_and_average_volume([{'radius': 3, 'height': 5}, {'radius': 4, 'height': 6}]) == (147.6548547187203, 73.82742735936014)
assert total_and_average_volume([{'radius': 1, 'height': 2}]) == (2.0943951023931953, 2.0943951023931953)
\end{lstlisting}

\subsection{An Example of BigCodeBench-Lite Pro }
\label{app:bcb-exp}
\begin{lstlisting}[caption= The outputs of GPT-4o on BigCodeBench-Lite Pro (BigCodeBench/355).,basicstyle=\ttfamily]
Traceback (most recent call last):
  File "evalpro/result/GPT-4o/bigcodebench_lite_pro/log/santized_results/case_20/gen_0.py", line 76, in <module>
    new_solution(points)
  File "evalpro/result/GPT-4o/bigcodebench_lite_pro/log/santized_results/case_20/gen_0.py", line 65, in new_solution
    voronoi_plot_2d(Voronoi(subset), ax=ax, show_vertices=False, line_colors=f'C{i}', point_size=2)
  File "_qhull.pyx", line 2677, in scipy.spatial._qhull.Voronoi.__init__
  File "_qhull.pyx", line 352, in scipy.spatial._qhull._Qhull.__init__
scipy.spatial._qhull.QhullError: QH6214 qhull input error: not enough points(1) to construct initial simplex (need 4)

\end{lstlisting}

\begin{lstlisting}[basicstyle=\ttfamily]
import numpy as np
from scipy.spatial import Voronoi, voronoi_plot_2d
import matplotlib.pyplot as plt
def task_func(points, seed=0):
    """
    Calculate the Voronoi diagram for a number of points in 2D and plot it.
    Note: this function will raise errors when input is invalid, for example wrong type or shape.
    Jittering is applied prior to plotting.

    Parameters:
    - points (np.ndarray): A numpy ndarray of shape (n_points, 2) with the coordinates of the points.
    - seed (int): Random seed for reproducibility. Defaults to 0.

    Returns:
    tuple (vor, ax): A tuple containing:
        - vor (Voronoi): A Voronoi object representing the Voronoi diagram of the points.
        - ax (Axes): The axes of the plotted Voronoi diagram.
    """
    if points.shape[1] != 2:
        raise ValueError("Input points should have shape (n_points, 2)")
        
    np.random.seed(seed)
    jittered_points = points + np.random.normal(scale=0.01, size=points.shape)

    vor = Voronoi(jittered_points)
    
    fig, ax = plt.subplots()
    voronoi_plot_2d(vor, ax=ax, show_vertices=False, line_colors='blue', point_size=2)
    plt.show()
    
    return vor, ax

def new_solution(points):
    """
    Given a set of points in 2D space, calculate and plot the Voronoi diagram for each subset of points obtained 
    by dividing the original set into three equal parts. Additionally, overlay the Voronoi diagrams of these subsets 
    on a single plot to visualize their combined effect.
    
    Parameters:
    - points (np.ndarray): A numpy ndarray of shape (n_points, 2) with the coordinates of the points.
    
    Returns:
    None
    """
    if len(points) < 3:
        raise ValueError("Need at least 3 points to divide into three subsets")
    
    # Sort points array for reproducibility
    points = np.array(sorted(points, key=lambda x: (x[0], x[1])))
    
    # Divide the points into three equal subsets
    n_points = len(points)
    part_size = n_points // 3
    
    subsets = [points[i*part_size:(i+1)*part_size] for i in range(3)]
    
    # Create a plot for overlay
    fig, ax = plt.subplots()

    # Plot the Voronoi for each subset
    for i, subset in enumerate(subsets):
        # Note: Here `task_func` is assumed to return the `ax`, which will be reused for plotting
        voronoi_plot_2d(Voronoi(subset), ax=ax, show_vertices=False, line_colors=f'C{i}', point_size=2)
        
    plt.title("Overlay of Voronoi Diagrams for the Three Subsets")
    plt.show()

# Test case 1: Basic test with 9 points
points = np.array([[0, 0], [0, 1], [1, 0], [1, 1], [2, 2], [2, 3], [3, 3], [3, 4], [4, 4]])
new_solution(points)

# Test case 2: Test with exactly 3 points
points = np.array([[0, 0], [1, 1], [2, 2]])
new_solution(points)

# Test case 3: Test with random points ensuring at least 9 points
points = np.random.rand(9, 2)
new_solution(points)
\end{lstlisting}

\newpage
\section{Error Statistics across Different Models}
\label{app:error-ana}
\begin{table*}[h!]
\centering

\label{tab:error}
\tiny
\begin{tabular}{l|l|cccccc|c}
\toprule
\multirow{2}{*}{\textbf{Model}}     & \multirow{2}{*}{\textbf{Dataset}} & \multicolumn{6}{c|}{\textbf{Error type}}                                                & \multirow{2}{*}{\textbf{All}} \\
                           &                          & AssertionError & NameError & ValueError & IndexError & TypeError & OtherError &                      \\

\midrule
O1-mini                    & HumanEval Pro            & 27             & 8         & 3          & 1          & 0         & 0          & 39  \\
                           & MBPP Pro                 & 89             & 15        & 6          & 2          & 4         & 4          & 120 \\
\rowcolor{lightblue}
                           & \textbf{All}                      & \textbf{116}            & \textbf{23}        & \textbf{9}          & \textbf{3}          & \textbf{4}         & \textbf{4}          & \textbf{159} \\
\midrule
GPT-4o                     & HumanEval Pro            & 28             & 11        & 2          & 1          & 0         & 0          & 41  \\
                           & MBPP Pro                 & 82             & 17        & 4          & 1          & 5         & 1          & 110 \\
\rowcolor{lightblue}
                           & \textbf{All}                      & \textbf{110}            & \textbf{28}        & \textbf{6}          & \textbf{2}          & \textbf{5}         & \textbf{1}          & \textbf{151} \\
\midrule
DeepseekCoder-V2-instruct  & HumanEval Pro            & 26             & 7         & 1          & 1          & 1         & 1          & 37  \\
                           & MBPP Pro                 & 79             & 12        & 4          & 3          & 7         & 3          & 108 \\
\rowcolor{lightblue}
                           & \textbf{All}                      & \textbf{105}            & \textbf{19}        & \textbf{5}          & \textbf{4}          & \textbf{8}         & \textbf{4}          & \textbf{145} \\
\midrule
DeepseekV2.5               & HumanEval Pro            & 30             & 8         & 2          & 1          & 2         & 0          & 43  \\
                           & MBPP Pro                 & 82             & 18        & 1          & 3          & 4         & 1          & 109 \\
\rowcolor{lightblue}
                           & \textbf{All}                      & \textbf{112}            & \textbf{26}        & \textbf{3}          & \textbf{4}          & \textbf{6}         & \textbf{1}          & \textbf{152} \\
\midrule
Qwen2.5-Coder-32B-instruct & HumanEval Pro            & 32             & 12        & 2          & 2          & 1         & 1          & 50  \\
                           & MBPP Pro                 & 89             & 16        & 3          & 1          & 4         & 1          & 114 \\
\rowcolor{lightblue}
                           & \textbf{All}                      & \textbf{121}            & \textbf{28}        & \textbf{5}          & \textbf{3}          & \textbf{5}         & \textbf{2}          & \textbf{164} \\
\midrule
Qwen2.5-Coder-7B-instruct  & HumanEval Pro            & 36             & 8         & 3          & 2          & 6         & 1          & 56  \\
                           & MBPP Pro                 & 93             & 14        & 3          & 3          & 18        & 2          & 133 \\
\rowcolor{lightblue}
                           & \textbf{All}                      & \textbf{129}            & \textbf{22}        & \textbf{6}          & \textbf{5}          & \textbf{24}        & \textbf{3}          & \textbf{189} \\
\midrule
Claude-3.5-sonnet & HumanEval Pro & 30  & 11 & 1  & 1 & 0  & 2  & 45  \\
                  & MBPP Pro      & 87  & 28 & 3  & 1 & 6  & 2  & 127 \\
\rowcolor{lightblue} & \textbf{All}  & \textbf{117} & \textbf{39} & \textbf{4}  & \textbf{2} & \textbf{6}  & \textbf{4}  & \textbf{172} \\
\midrule
LLaMa-3-70B-instruct & HumanEval Pro & 44  & 10 & 3  & 2 & 2  & 4  & 65  \\
                     & MBPP Pro      & 100 & 12 & 2  & 2 & 14 & 8  & 138 \\
\rowcolor{lightblue} & \textbf{All}  & \textbf{144} & \textbf{22} & \textbf{5}  & \textbf{4} & \textbf{16} & \textbf{12} & \textbf{203} \\
\midrule
Codestral-22B        & HumanEval Pro & 45  & 13 & 3  & 3 & 2  & 1  & 67  \\
                     & MBPP Pro      & 102 & 16 & 3  & 1 & 12 & 3  & 137 \\
\rowcolor{lightblue} & \textbf{All}  & \textbf{147} & \textbf{29} & \textbf{6}  & \textbf{4} & \textbf{14} & \textbf{4}  & \textbf{204} \\
\midrule
OpenCoder-8B-base    & HumanEval Pro & 47  & 43 & 0  & 3 & 5  & 2  & 100 \\
                     & MBPP Pro      & 114 & 43 & 2  & 2 & 14 & 6  & 181 \\
\rowcolor{lightblue} & \textbf{All}  & \textbf{161} & \textbf{86} & \textbf{2}  & \textbf{5} & \textbf{19} & \textbf{8}  & \textbf{281} \\
\midrule
OpenCoder-8B-instruct & HumanEval Pro & 42  & 15 & 2  & 1 & 5  & 2  & 67  \\
                      & MBPP Pro      & 118 & 22 & 3  & 1 & 11 & 4  & 159 \\
\rowcolor{lightblue}  & \textbf{All}  & \textbf{160} & \textbf{37} & \textbf{5}  & \textbf{2} & \textbf{16} & \textbf{6}  & \textbf{226} \\
\midrule
Qwen2.5Coder-1.5B-base & HumanEval Pro & 56  & 25 & 7  & 1 & 9  & 5  & 103 \\
                       & MBPP Pro      & 117 & 37 & 3  & 4 & 14 & 21 & 196 \\
\rowcolor{lightblue}   & \textbf{All}  & \textbf{173} & \textbf{62} & \textbf{10} & \textbf{5} & \textbf{23} & \textbf{26} & \textbf{299} \\
\midrule
Qwen2.5Coder-7B-base   & HumanEval Pro & 45  & 15 & 3  & 4 & 5  & 2  & 74  \\
                       & MBPP Pro      & 99  & 21 & 1  & 3 & 16 & 6  & 146 \\
\rowcolor{lightblue}   & \textbf{All}  & \textbf{144} & \textbf{36} & \textbf{4}  & \textbf{7} & \textbf{21} & \textbf{8}  & \textbf{220} \\
\midrule
Qwen2.5Coder-32B-base  & HumanEval Pro & 39  & 15 & 3  & 3 & 1  & 2  & 63  \\
                       & MBPP Pro      & 90  & 17 & 2  & 2 & 7  & 4  & 122 \\
\rowcolor{lightblue}   & \textbf{All}  & \textbf{129} & \textbf{32} & \textbf{5}  & \textbf{5} & \textbf{8}  & \textbf{6}  & \textbf{185} \\
\midrule
Yi-Coder-9B            & HumanEval Pro & 48  & 31 & 2  & 5 & 3  & 5  & 94  \\
                       & MBPP Pro      & 92  & 37 & 1  & 3 & 12 & 5  & 150 \\
\rowcolor{lightblue}   & \textbf{All}  & \textbf{140} & \textbf{68} & \textbf{3}  & \textbf{8} & \textbf{15} & \textbf{10} & \textbf{244} \\
\midrule
Yi-Coder-9B-Chat       & HumanEval Pro & 47  & 12 & 1  & 3 & 3  & 0  & 66  \\
                       & MBPP Pro      & 96  & 19 & 1  & 2 & 11 & 4  & 133 \\
\rowcolor{lightblue}   & \textbf{All}  & \textbf{143} & \textbf{31} & \textbf{2}  & \textbf{5} & \textbf{14} & \textbf{4}  & \textbf{199} \\
\midrule
GPT-4-Turbo            & HumanEval Pro & 33  & 8  & 3  & 1 & 1  & 0  & 46  \\
                       & MBPP Pro      & 91  & 18 & 1  & 1 & 5  & 0  & 116 \\
\rowcolor{lightblue}   & \textbf{All}  & \textbf{124} & \textbf{26} & \textbf{4}  & \textbf{2} & \textbf{6}  & \textbf{0}  & \textbf{162} \\
\midrule
DeepseekCoder-33B-base & HumanEval Pro & 55  & 16 & 2  & 2 & 3  & 5  & 83  \\
                       & MBPP Pro      & 108 & 23 & 5  & 1 & 8  & 10 & 155 \\
\rowcolor{lightblue}   & \textbf{All}  & \textbf{163} & \textbf{39} & \textbf{7}  & \textbf{3} & \textbf{11} & \textbf{15} & \textbf{238} \\
\midrule
DeepseekCoder-33B-instruct & HumanEval Pro & 49  & 14 & 2  & 2 & 4  & 0  & 71  \\
                          & MBPP Pro      & 101 & 16 & 2  & 1 & 10 & 6  & 136 \\
\rowcolor{lightblue}      & \textbf{All}  & \textbf{150} & \textbf{30} & \textbf{4}  & \textbf{3} & \textbf{14} & \textbf{6}  & \textbf{207} \\
\midrule
DeepseekCoder-6.7B-base & HumanEval Pro & 59  & 24 & 4  & 4 & 6  & 9  & 106 \\
                        & MBPP Pro      & 128 & 25 & 3  & 3 & 14 & 14 & 187 \\
\rowcolor{lightblue}    & \textbf{All}  & \textbf{187} & \textbf{49} & \textbf{7}  & \textbf{7} & \textbf{20} & \textbf{23} & \textbf{293} \\
\midrule
DeepseekCoder-6.7B-instruct & HumanEval Pro & 46  & 15 & 4  & 4 & 2  & 2  & 73  \\
                           & MBPP Pro      & 107 & 30 & 4  & 2 & 17 & 2  & 162 \\
\rowcolor{lightblue}       & \textbf{All}  & \textbf{153} & \textbf{45} & \textbf{8}  & \textbf{6} & \textbf{19} & \textbf{4}  & \textbf{235} \\
\midrule
Magicoder-S-DS          & HumanEval Pro & 49  & 11 & 6  & 4 & 5  & 0  & 75  \\
                        & MBPP Pro      & 107 & 21 & 2  & 2 & 20 & 4  & 156 \\
\rowcolor{lightblue}    & \textbf{All}  & \textbf{156} & \textbf{32} & \textbf{8}  & \textbf{6} & \textbf{25} & \textbf{4}  & \textbf{231} \\
\midrule
WaveCoder-Ultra-6.7B    & HumanEval Pro & 51  & 12 & 2  & 3 & 4  & 2  & 74  \\
                        & MBPP Pro      & 113 & 20 & 2  & 4 & 8  & 4  & 151 \\
\rowcolor{lightblue}    & \textbf{All}  & \textbf{164} & \textbf{32} & \textbf{4}  & \textbf{7} & \textbf{12} & \textbf{6} & \textbf{225} \\

\bottomrule
\end{tabular}
\caption{Error type of Different Models on HumanEval Pro and MBPP Pro.}
\end{table*}